# Epitaxy of Semiconductor–Superconductor nanowires

P. Krogstrup[1,*], N. L. B. Ziino[1], W. Chang[1], S. M. Albrecht[1], M. H. Madsen[1], E. Johnson[1,2], J. Nygård[1,3], C. M. Marcus[1] and T. S. Jespersen[1,*]

[1] Center for Quantum Devices, Niels Bohr Institute, University of Copenhagen, Denmark
[2] Department of Wind Energy, Technical University of Denmark, Risø Campus, Roskilde, Denmark
[3] Nano-Science Center, Niels Bohr Institute, University of Copenhagen, Denmark

**Controlling the properties of semiconductor/metal interfaces is a powerful method for designing functionality and improving the performance of electrical devices. Recently semiconductor/superconductor hybrids have appeared as an important example where the atomic scale uniformity of the interface plays a key role for the quality of the induced superconducting gap. Here we present epitaxial growth of semiconductor-metal core-shell nanowires by molecular beam epitaxy, a method that provides a conceptually new route to controlled electrical contacting of nanostructures and for designing devices for specialized applications such as topological and gate-controlled superconducting electronics. Our materials of choice, InAs/Al, are grown with epitaxially matched single plane interfaces, and alternative semiconductor/metal combinations allowing epitaxial interface matching in nanowires are discussed. We formulate the grain growth kinetics of the metal phase in general terms of continuum parameters and bicrystal symmetries. The method realizes the ultimate limit of uniform interfaces and appears to solve the soft-gap problem in superconducting hybrid structures.**

\* Corresponding Authors: krogstrup@nbi.dk, tsand@nbi.dk

The discovery of materials with topological classification is interesting as fundamental physics, and also holds promise as the basis for quantum information processing based on manipulations, i.e. braiding, of quasiparticles that exist at the boundaries of different topologies. A key advance in the field was the realization that such quasiparticles, Majorana modes, can be formed relatively straightforwardly in semiconductor nanowires (NWs) coupled to conventional superconductors[1,2]. This realization was in turn followed up by a number of experiments showing strong evidence for Majorana modes[3,4,5,6]. So far, however, all realizations fabricated by conventional methods show unexpected low-energy states below the proximity-induced superconducting gap. Such *soft-gap* states are a source of decoherence of Majorana modes, and thus detrimental to topological quantum information processing. Recently, theory has implicated disorder in the semiconductor/superconductor (SE/SU) interface as the source of the soft gap[7,8]. This is a familiar situation in the history of semiconductor technology, where device performance has increased together with interface quality, leading ultimately to semiconductor hetero-epitaxy and bandgap engineering[9]. Recent progress in the formation of both axial and radial heterostructure semiconductor nanowires[10,11,12], has resulted in devices with new and exciting functionalities[13,14,15]. However, even though epitaxial interfaces constitute the ultimate limit of uniformity, epitaxy of metals[16] (and superconductors, in particular) has so far not been combined with the world of semiconductor nanowire epitaxy.

In this work, we introduce a method to grow epitaxial SE/SU InAs/Al nanowire heterostructures in a two-step process in-situ by molecular beam epitaxy (MBE). The results are clean epitaxial SE/SU interfaces and uniform crystal morphologies. The growth of Al is analysed in terms of the general control parameters and bicrystal symmetries. The analysis thus applies for arbitrary material combinations, and we discuss other possible well-matched SE/SU systems. Finally, the InAs/Al contact resistance and superconducting properties is characterized at low temperatures. The growth method presented here not only solves the soft-gap problem described above, making the hybrid crystals promising candidates for topological quantum devices, but also provide a general and conceptually new approach for controlled contacting to NWs.

**InAs/Al semiconductor/superconductor NW epitaxy**

InAs/Al is a particularly interesting materials combination for topological SE/SU quantum devices – InAs has a strong spin orbit coupling and a large *g*-factor, and Al has a long superconducting coherence length and is compatible with standard fabrication techniques. In this work, we study InAs NWs grown either along the <0001>B or the <01-10> direction, allowing a study of the Al phase formation on two different types of InAs facets. Typical NW lengths are 5-10 μm and diameters are 60-100nm. The "conventional" <0001> NWs are grown on [111]B InAs substrates using either patterned (Fig. 1a) or randomly distributed Au particles. The resulting NWs have six {1-100} side facets on which Al was subsequently grown either while rotating the substrate resulting in Al on all facets, or with the substrate orientation fixed either for growth on two (Fig. 1) or on three facets. Kinked NWs with <01-10>$_{WZ}$ (or <11-2>$_{ZB}$) orientations were grown epitaxially from the [0001] stem[17], see supplementary information S5. The structures have a rectangular-like cross section and are WZ dominated, but the kinking process typically induces stacking faults and even ZB segments.

The top facets towards the aluminum ($[0001]B_{WZ}$ or $[111]B_{ZB}$) have, however, identical single plane surfaces for the ZB and WZ structures.

Figure 1b shows an example of a <0001> InAs/Al hybrid with 10nm Al grown on two facets and Fig. 1c shows a high resolution transmission electron microscope (TEM) image of the InAs/Al interface, already demonstrating one of the main results of this work: the growth of completely uniform, oxide-free and atomically abrupt InAs/Al interfaces. In Fig. 1c, the atomic planes of both crystal phases can be seen, allowing analysis of the detailed epitaxial relations at the interfaces. In general, when a lattice mismatch of a given material combination is large, interfacial domains consisting of $n_{SE}$ and $n_{SU}$ interfacial units of *SE* and *SU* form in order to reduce the interfacial energy[18]. Thus, to specify a *SE/SU* interface with a given out of plane orientation, we consider the interfacial atoms which contribute with broken bonds in the case of the corresponding ideal flat surfaces, and use the following compact notation for the in-plane interfacial domain: $\left(\frac{n_{SU,\parallel}}{n_{SE,\parallel}}, \varepsilon_\parallel\right) \times \left(\frac{n_{SU,\perp}}{n_{SE,\perp}}, \varepsilon_\perp\right)$. Here $n_{SU}/n_{SE}$ and $\varepsilon_i$ denote the ratio of units in the domain and the residual mismatch of the relaxed structures, along the in-plane directions parallel and transverse to the NW axis, respectively.

Figures 2a, 2d, and 2f show high-resolution TEM images of the interfaces for the structures we generally observe, corresponding to three different categories of InAs/Al hybrids: the <0001> InAs NWs with thin <~10 nm Al (Fig. 2a), <0001> NWs with >~ 30nm Al shells (Fig. 2d), and <01-10> NWs with Al on the (111)B facet (Fig. 2f), respectively. In each case, the crystal orientations are indicated on the figures.

To assign the bicrystal match, also the overall morphology of the Al shells gives information about the out-of plane orientation, as illustrated in the TEM images of Fig. 3 for three (two) different Al phase thicknesses on <0001> (<01-10>) InAs NWs. For the thinnest Al on <0001> NWs and for the <01-10> NWs the free surface of the Al is parallel to the NW axis, indicating that the Al has the <111> direction normal to the surface. For the thicker Al on the conventional <0001> NWs, however, the Al develops a clear grain structure and the surface becomes faceted (see below). Also the overall bending of the hybrids provides useful information. The bending can originate from either interface mismatch or differences in the thermal expansion coefficients for InAs and Al. The latter would result in hybrids always bending away from the Al when the substrate is elevated to room temperature after growth. This is not observed, suggesting a bending primarily due to the residual strain related to the formation of interfacial domains. In the half-shell geometry, the Al usually causes the hybrids to bend either towards or away from the Al-covered side, depending on the orientational relationship of the components, which provides indirect information about the interface mismatch. As an example, Fig.

3f shows an SEM image for the case of a NW with 10 nm Al grown on two facets. The hybrid bends towards the Al and thus suggests an interface with compressive (tensile) strain along the NW axis in the InAs (Al).

Consider first the InAs/Al interface for the thin Al shell on two facets (Fig. 2a), where the Al has low energy (111) planes normal to the NW facets and thus attains a uniform single facetted morphology (Fig. 3a). For the domain formation, we consider two candidates in this orientation: either a small domain $\left(\frac{3_{[11\bar{2}]}}{2_{[0001]}}, 6.5\%\right) \times \left(\frac{3_{[1\bar{1}0]}}{2_{[11\bar{2}0]}}, 0.3\%\right)$ with a large mismatch along the NW axis, or a larger domain $\left(\frac{7_{[11\bar{2}]}}{5_{[0001]}}, -0.5\%\right) \times \left(\frac{3_{[1\bar{1}0]}}{2_{[11\bar{2}0]}}, 0.3\%\right)$ with a smaller mismatch. The two cases are simulated in Fig. 2b and c, respectively. As shown in Fig. 3f, the NW bends towards the Al, suggesting the formation of the larger domain with the negative and much smaller mismatch.

For the kinked <0-110> InAs NWs, the Al forms with the <111> out-of-plane orientation with uniform morphologies as a result (Fig 3 d,e). Because of the different InAs facets this orientation has the possibility of forming a highly ordered epitaxial domain match, $\left(\frac{3_{[11\bar{2}]}}{2_{[11\bar{2}]_{ZB}}}, 0.3\%\right) \times \left(\frac{3_{[1\bar{1}0]}}{2_{[1\bar{1}0]_{ZB}}}, 0.3\%\right)$ while keeping the low surface energy (111) out-of-plane orientation. The slightly positive strain along the NW length of this match is consistent with an observed bending away from the Al (see supplementary information S5). Growing thicker shells on the kinked wires does not seem to change the preferred crystal orientations (Fig. 3e). Surprisingly, however, for thicker shells grown on the <0001> InAs NWs (Fig. 2d and Fig. 3c), the dominating crystal orientation of the Al appears to change from the <111> to the <11-2> out of plane orientation. With this orientation the Al can form small and remarkably well matched $\left(\frac{1_{[111]}}{1_{[0001]}}, 0.3\%\right) \times \left(\frac{3_{[1\bar{1}0]}}{2_{[11\bar{2}0]}}, 0.3\%\right)$ domains, as shown in Fig. 2e. Since, the crystal transition from <111> to <11-2> out of plane orientation takes place gradually, the NW curvature stems from the initial <111> out of plane domain match. As the low-energy (111) planes of Al are not parallel to the wire axis with the <11-2> orientation, the overall morphology becomes faceted as seen in Fig. 3c.

To understand the growth mechanisms in more detail, and in particular the transition from <111> to <11-2> out of plane orientation for the Al shells on the <0001> InAs NWs, we use a theoretical continuum formalism for growth kinetics[19] to explain the metallic phase formation on semiconductor NW facets. While the details of the model are discussed in supplementary information S1 and S2, the qualitative results are as follows: In the beginning of the Al growth where the Al thickness is small, the out of plane orientation of the SU phase is mainly determined by surface energy minimization, while the in-plane orientation is determined by the interface energy minimization. This is consistent with the orientation in Fig. 2a and Fig. 3a of the thin shells having the low energy (111) planes parallel to the NW resulting in a planar morphology with a low surface energy, and still having distinct in-plane relations giving by high bicrystal symmetries (see supplementary information S3).

As the metallic phase grows thicker the surface-to-volume ratio decreases, and the stress induced from the InAs/Al interface and maybe more importantly from incoherent boundaries of Al grains meeting on a side facet or across adjacent side facets, provide increasingly strong driving forces for reconstruction into a less strained and lower total energy configuration. The <111> to <11-2> transition is indeed consistent with the ability of the <11-2> orientation to form large coherent single crystals while maintaining simultaneous epitaxial match on all facets. This remarkable situation is illustrated in Fig. 4 and relies on the bicrystal symmetries in three dimensions. In general, for a given out of plane orientation there exist a number of distinguishable grains with indistinguishable interfacial domains if the order of the in-plane rotational symmetries of SE and SU are different from the corresponding bulk rotational symmetries (see supplementary information S3). All types of interfaces in Fig. 2 have a two-fold in-plane degeneracy, meaning that two grain variants, denoted $\alpha$ and $\beta$, can form with equal probability. Figure 4a shows a <0001> InAs/Al full-shell hybrid where two different orientations of the Al is clearly seen along the nanowire. However, we only rarely see the grain orientation change around the NW (Fig. 4e). For the <11-2> FCC orientation, the WZ and FCC crystals have a six-fold and three-fold rotational symmetry along the NW axis, respectively. Thus, two FCC grains of the same variant on adjacent facets will meet in an incoherency across the <01-10> facets of the WZ. However, if the variants alternate around the NW, it will form a single crystal with <01-10>/<11-2> type domains on all six side facets as shown in Fig. 4d. Only because the 6-fold rotational symmetry of the WZ NW axis is really a screw axis, the perfect bicrystal symmetry is broken (Fig. 4f) and this may induce a small strain field at the edge where the side facets meets. Thus, if identical variants nucleate on adjacent facets, the high excess free energies of the incoherent boundaries will drive an elimination/reconstruction of the smaller on expense of the larger. An example of a kinetically locked incoherent interface is shown in Fig. 4e.

A different situation occurs for grain boundaries meeting along the direction of the nanowire: $\alpha$ and $\beta$ grains form low energy coherent twin boundaries, as simulated in Fig. 4f, which do not lead to a driving force sufficiently strong to eliminate the grain boundaries. The resulting structure is thus a shell that is coherent around the circumference, but with alternating single plane variants along the nanowire as seen in Fig. 4a. We do not observe a structural transition as a function Al thickness on the kinked NWs because the Al are already in symmetry with InAs at thin phases.

**Electrical properties of Al/InAs hybrids**

Having established the existence of highly ordered epitaxial SE/SU interfaces, we next consider the electrical transport properties of the materials and interfaces. To investigate the superconducting properties of the Al film, four-terminal devices were fabricated on a 90nm diameter nanowire with a 13nm full shell as illustrated in Fig. 5a and b. The inset to Fig. 5c shows the temperature dependence of the four-terminal resistance. The Al shell has a normal state resistance of ~50Ω for temperatures above ~1.6K and shows a gradual transition to the superconducting zero-resistance state which is fully developed by 1.3K. The in-plane critical magnetic field $B_c$ and the coherence length $\xi(0)$ of the shell are important parameters related to the quality of the Al film. The parallel magnetic field dependence in Fig. 5c shows a zero-resistance state persisting up to 1.3 Tesla, only interrupted by finite resistance peaks appearing for fields corresponding to an odd multiple of half flux quanta threading the wire. The periodic resistance peaks are specific to the full-shell geometry, and termed the destructive regime of the Little-Parks effect in a cylindrical superconductor.[20,21] Importantly, the appearance of the destructive regime directly shows that the coherence length $\xi(0)$ exceeds the diameter of the cylinder, d =~ 100nm. Only recently and using specialized techniques, have such long coherence lengths been realized in micro-fabricated aluminum nanoscale structures[22] and Fig. 5c, thus confirm the high quality of our MBE grown Al contact shell. The Little-Parks oscillations in Fig. 5c are visible up to the critical magnetic field $B_c$ ~1.9T where the superconducting state is destroyed. This value is consistent with critical fields reported for 13 nm high quality planar Al films[23]. Note that a topological phase requires critical fields exceeding $2\Delta^*/g\mu_B$ where $\Delta^*$ is the induced gap, $g$ the $g$-factor of InAs and $\mu_B$ the Bohr magneton. For InAs and Al, typical values are $\Delta^* = 190$ μeV and $g=10$, and $2\Delta^*/g\mu_B \sim 0.7$T well below $B_c$.

The electrical properties of the epitaxial SE/SU interface were studied in devices where the semiconductor core was exposed by selectively etching a segment of the Al (Fig. 5d, inset). The device acts as a nanowire field effect transistor with the epitaxial shell serving as contacts, and Fig. 5d shows the measured conductance as a function of the voltage $V_g$ applied to the conducting back plane for various temperatures above $T_c$. As is generally observed, the nominally undoped InAs nanowire acts as an $n$-type semiconductor. It is depleted at $V_g = -10$V and the conductance increases with $V_g$ to 2.8 $e^2/h$ at $V_g = 10$V. Other devices with shorter exposed InAs segments had conductivities up to 6 $e^2/h$ at $V_g = 10$ V. These values are comparable to the best results we have achieved for devices of comparable lengths and diameters using conventional $(NH_4)_2S_x$ passivation or argon milling for removing the native InAs oxide prior to metal evaporation. This indicates that the epitaxial shell forms a barrier-free metal/semiconductor contact as is further supported by the temperature dependence of the transfer curves in Fig. 5d: For $V_g > 2$ V the conductance increases upon cooling due to the reduction of phonon scattering rather than decreases as is most often observed for imperfect contacts due to the reduction of thermally excited transport over contact barriers.

To study the quality of the superconducting gap induced in the InAs core, devices with one normal electrode were fabricated (Fig. 5e, top schematic) allowing tunnel spectroscopy of the density of states of the semiconductor core when operated at gate voltages close to pinch-off. A typical measurement of the differential conductance as a function of bias is shown in Fig. 5e. The induced gap $\Delta^*$ ~190μeV is close to the value for bulk aluminum, and for sub-gap bias voltages ($V_{sd} \ll \Delta^*/e$), the conductance is limited by the noise-floor in our measurements, and at least reduced by two orders of magnitude compared to the normal state (above gap) value. This should be compared to reduction of a factor of ~5 which is state-of-the-art for evaporated superconducting contacts[3,4,5] and also the best value we have achieved with evaporated Al contacts on the same InAs nanowires. The hard gap is further analyzed in Ref. [24].

While the TEM analysis in Fig. 1-4 establishes the epitaxial properties of the InAs/Al interfaces, the results in Fig. 5 demonstrate that devices with epitaxial contacts can be fabricated by standard fabrication techniques maintaining the high quality of the superconducting shell. Importantly, the results confirm that the epitaxial interface provides low resistance superconductor-semiconductor contacts, and that the epitaxial shell induces a hard superconducting gap in the InAs core.

**Engineering the interface**

In addition to the epitaxial and uniform contact interfaces, which cannot be obtained by conventional fabrication techniques, our method also allows engineering of contacts and interfaces using the flexibility and control of the MBE technique. This opens new possibilities and is desirable for a number of applications. For example, intermediate tunnel barriers have been suggested[7] to enhance the induced proximity gap and minimize quasiparticle poisoning[25]. Also the technique could be used to grow superconducting contact heterostructures with built-in normal metal quasi-particle traps.[26] In the growth of semiconductor heterostructures, barriers of high band-gap materials are widely used, and in Fig. 6 we demonstrate this method with an InAs/Al half shell hybrid with a 3 nm segment of high band-gap AlAs (band gap of 2.12 eV) grown to separate the Al from the InAs. Since AlAs is not lattice matched with InAs, a specialized growth sequence was developed to grow the AlAs in its relaxed form and thus avoid excessive strain and strong bending of the NW (see Supplementary section 6).

**Alternative material combinations**

It should be emphasized that the growth of hybrid metal-NW crystals is not restricted to the InAs/Al system. Due to the large number of possible orientations, however, it is challenging not only to predict which material combinations can be matched but also which materials can be tuned in terms of growth kinetics to form uniform NW heterostructures. As a first step, we have searched for combinations of cubic metals which match InAs, InSb, and GaAs semiconductor nanowires grown in the conventional <111>$_{ZB}$ (or the similar <0001>$_{WZ}$) directions. As described above, the symmetries of these orientations are particularly appealing if aligned along the same class of symmetries, as they may allow single coherent grains to match the semiconductor across multiple facets. Thus we expect these orientations to be likely to occur in the thick film limit (grain boundary driven regime) if they are matched in a given SE/SU system. In the Supplementary section 4 we list the expected strain for a large number of metals. In addition to the InAs/Al system (2/3 domain ratio, 0.3% strain), other noteworthy well-matched combinations include InAs/Au (2/3 domain ratio, 1.0% strain) which may serve as model contact materials for non-superconducting applications, and InAs/V (1/2 domain ratio, 0.3% strain) and InSb/Nb (1/2 domain ratio 1.8% strain) which are important combinations for high-critical temperature and magnetic field superconducting contacts to strong spin-orbit semiconductors.

**Conclusions**

In conclusion, we have developed nanowire heterostructures consisting of InAs and Al layers, grown by MBE. We achieve highly ordered epitaxial interfaces with only 0.3% mismatch while maintaining crystallinity of both the InAs and Al components. We have also developed a general formalism to analyze growth kinetics of the metallic phase formation and the epitaxial properties of the semiconductor/superconductor interface bilayers. Growth methods were developed for achieving both fully covering and partly covering Al shells. Both the experiment and the model indicate that the epitaxial properties can be maintained around the full circumference of the InAs core. We discussed the use of bandgap engineering to control the interface properties and showed, as an example, the inclusion of controlled large band-gap AlAs barrier at the interface between InAs and Al. We considered the compatibility of the technique for other material combinations and show that well matched interfaces are also possible for InAs/Au and InSb/Nb, which constitute model systems for normal metal/semiconductor and high critical temperature and magnetic field superconductor/semiconductor hybrids, respectively.

InAs/Al devices were fabricated and characterized electrically at low temperature, confirming the high quality of the epitaxial Al and the Al/InAs interface. For temperatures below the superconducting transition temperature, Al induced a superconducting gap into the InAs by virtue of the proximity effect[27,28,29]. In contrast to previous studies, the induced gap remains *hard* i.e., free of sub-gap states, likely due to the high quality of the Al shell and the perfectly uniform InAs/Al interface. These hybrid structures thereby remove one of the main obstacles for using semiconductor nanowires as the backbone in future schemes of topological quantum information based on Majorana Fermions[7,8]. In this context we note also that the InAs/Al epitaxial hybrids fulfill all basic requirements remaining for use in Majorana devices: strong spin-orbit coupling, large critical parallel magnetic field, and gate-tunability. Thus we believe that the approach developed in this paper is relevant for all device applications of semiconductor nanowires and, in particular, crucial for the future developments of the topological quantum information technology.

**Methods**

The InAs nanowires were grown in two different crystal directions on (111)B InAs substrates by the Au-catalyzed vapor-liquid-solid method in a solid-source Varian GEN-II MBE system. The first type is the conventional NWs with an axial (0001)B wurtzite orientation growing vertical on the substrate, with a corresponding planar growth rate of 0.5µm/hr and a V/III ratio of ~100 for 30 minutes at a substrate temperature of 420 C. These conditions give us a pure WZ crystal structure with flat {1-100} side facets. The second type is first grown like the conventional NWs, but after a certain time the wire growth direction was kinked into one of the six equivalent <1-100> orientations. This was obtained by either a short flush of Ga or a short decrease in temperature to make the growth region unstable.[19] Following InAs growth, the substrate temperature is kept at 200 C until the background pressure in the growth chamber is below $10^{-10}$ Torr. In our chamber this takes about three hours. This is an important step to avoid any material sticking at the NW surface before the metal growth. Hereafter, the substrate is cooled below -30 C, by turning of all power sources that can act as heat sources (power supply for substrate holder, ion gauges, light sources). This process typically takes more than 8 hours in our chamber. For the half shell growth the substrate rotation is disabled and the substrate is visually oriented with an accuracy of ~2-3 degrees to have the desired crystal orientation facing the Al cell (RHEED can be used as an alternative for substrate alignment). The thickness of the metal phase on the NW facets is given by $s(t) = \chi \Omega_S \sin(\varphi) f \cdot t$, where $\Omega_S$ is the atomic volume, $\varphi$ is the angle of the incoming beam with respect to the facet normal, $f$ the incoming flux of growth species, and $\chi$ is a correction factor for the time the beam is effectively hitting the facets. The corresponding planar growth rate ($\Omega_S f \cos(\varphi)$) for the Al growth was 0.3-0.5µm/hr. During deposition, the substrate temperature as measured on the thermocouple increases about 1-2 degrees before reaching a steady state temperature only a few minutes after the growth start. After growth the substrate is turned away from the sources and left in the buffer chamber at room temperature, before any heat are turned on in the growth chamber.

All structural simulations of the NW crystals in Fig.2 and Fig.4 are done using the software program Vesta[30].

Electrical devices were fabricated as follows: The InAs/Al hybrid nanowires were liberated from their growth substrate by a brief sonication in methanol, and a small amount of the resulting suspension was deposited on doped Si substrates that were capped with 500 nm of SiO$_2$. Wires were located with respect to predefined alignment marks using optical dark field microscopy and the ends of the wires were contacted using electron beam lithography (6% copolymer, 4% poly-[methyl methacrylate] (PMMA)) and electron beam evaporation of ~5/100 nm of Ni/Au or Ti/Au (AJA International, ATC ORION evaporator). The oxide on the Al surface was removed by 120 s of Kaufmann argon ion milling performed inside the metal evaporation chamber (300 V acceleration, 46 mA emission current). This procedure reproducibly created contact to the Al shell. For the devices with exposed InAs, narrow etch windows were defined in 4% PMMA resist by e-beam lithography, and the shell was removed by a ~2 s etch in 12% hydrofluoric acid. Finally, the device is coated in 20-30 nm of hafnium oxide using atomic layer deposition.

Low temperature electrical measurements were performed in a dilution refrigerator using standard lock-in techniques with a 10 µV ac excitation.

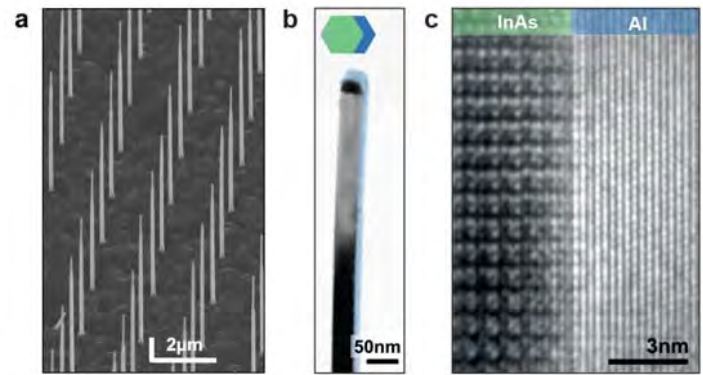

**Fig. 1. Overview of epitaxial InAs/Al hybrids.** **a** Tilt-view scanning electron micrograph of an array of epitaxial InAs/Al NWs grown on an InAs (111)B substrate. **b** TEM micrograph of the top part of a NW taken from the sample shown in **a** with the Al shell highlighted in blue. The Al is ~8nm thick and covers two facets of the wire, as illustrated in the schematic cross section (inset). The high resolution TEM image in panel **c** shows that the Al forms a perfectly sharp and uniform interface to the InAs core. In this example, the InAs core was grown in the $[0001]_{WZ}$ direction and the crystal orientation of the Al along the whole length of the NW is with the high symmetry and low energy [111] orientation normal to the interface.

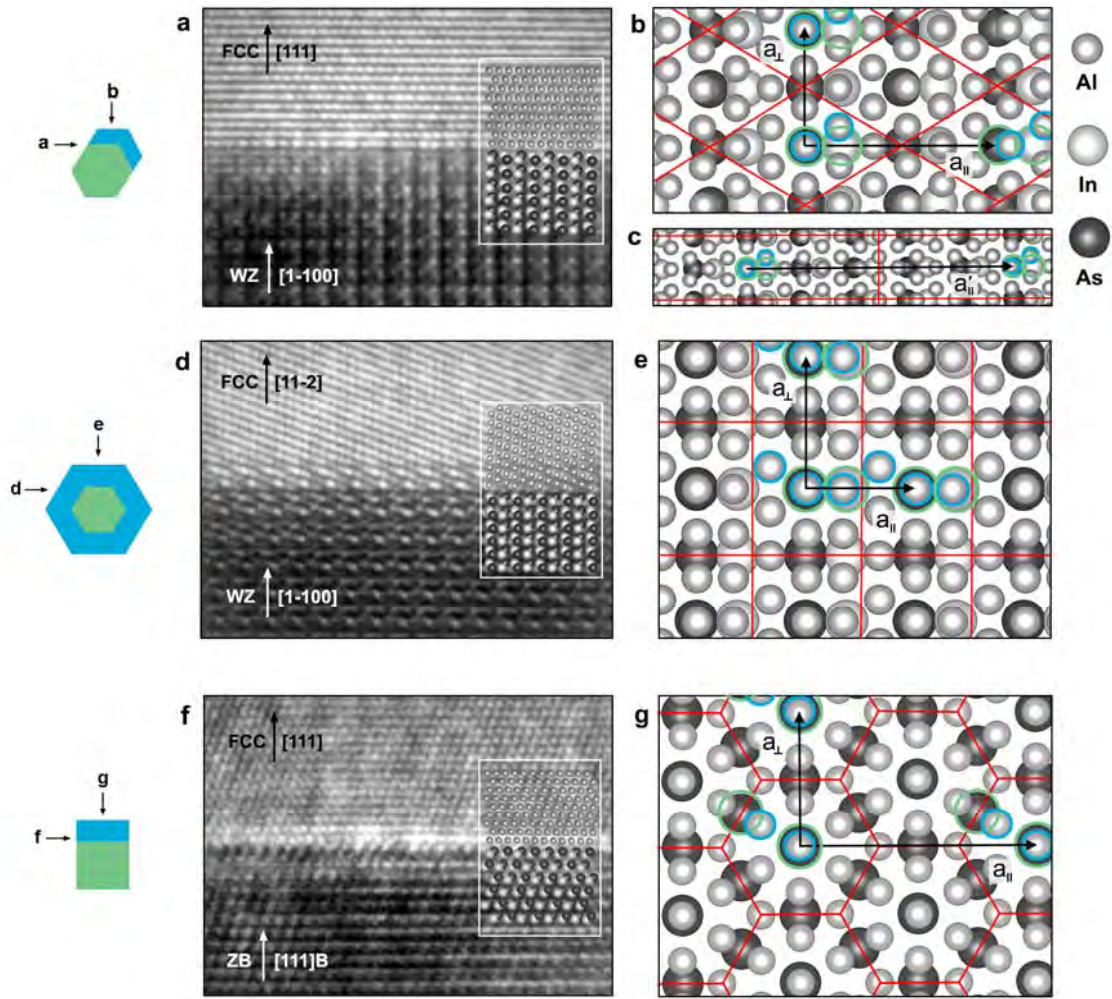

**Fig. 2. Domain-matched InAs/Al interfaces.** Three dominant epitaxial bicrystal matches formed in three types of NW hybrids. (**a**, **d**, **f**) TEM images along InAs/Al interface, along with simulations (insets). The viewing orientations are illustrated by the leftmost diagrams. (**b**, **c**, **e**, **g**) Top view on corresponding single plane interface structures for the bicrystal orientations, as if the InAs and Al phases are relaxed (lattice constants are taken from bulk values). Red lines indicate primitive domains assuming a perfect domain match, and the highlighted circles (blue and green) specify the atoms in one interfacial unit of each constituent in the parallel and transverse directions as shown with the vectors. An atomic displacement with respect to the circles illustrates the domain mismatch. Cubic notation is used in **f**, because the top-part of the [1-100] wire has adopted the cubic ZB structure. See text for discussions on the growth kinetics behind the domain formation.

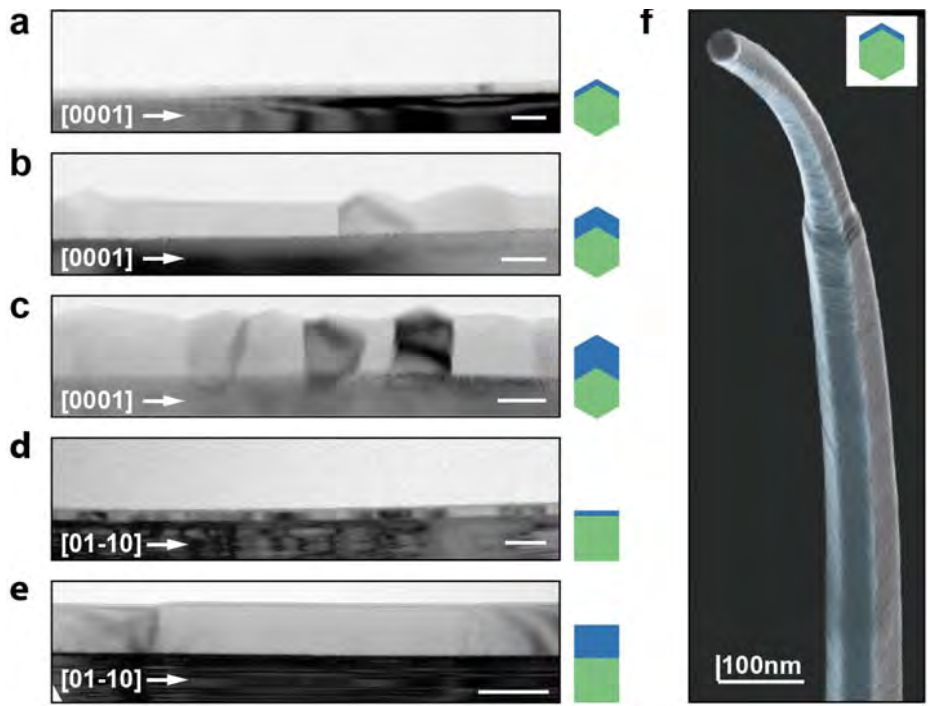

**Fig. 3. Al layer morphology and asymmetric strain.**

Transmission electron micrographs of different types of Al/InAs hybrid nanowires with varying Al thicknesses. **a-c** InAs grown along the [0001], where a structural transition of the Al with out of plane orientation from {111} (planar surface) to {11-2} (faceted surface) is observed beyond a critical thickness of roughly $r_{SU} \sim 20$ nm. **d-e** InAs NWs grown along the <1-100> WZ structure with stacking faults along the length, and the Al shell remains uniform with {111} out of plane orientations. Scale-bars in **a-e** are 25 nm. The diagrams on the right illustrates the corresponding NW cross-section of the InAs(green)/Al(blue) hybrids. **f** Tilt view SEM close-up of a type-1 half-shell NW demonstrating the asymmetric strain which causes the NW to bend towards the Al due to the InAs/Al interfacial domain mismatch.

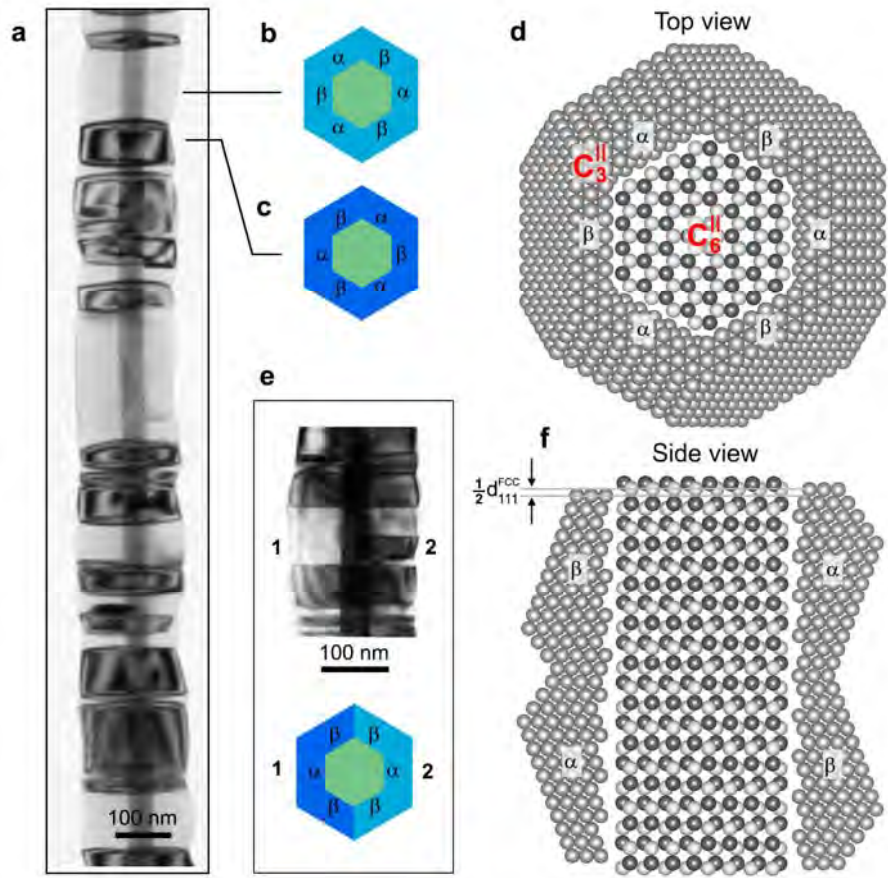

**Fig. 4. Full shell epitaxial bicrystal match.** a TEM image of a $[0001]_{WZ}$ InAs NW (seen as the dark core) with a thick layer of epitaxial Al grown on all six {1-100} facets. The dominating Al grain orientation is of the {11-2} type. The two grain orientations, $\alpha$ and $\beta$, are clearly distinguishable by the TEM diffraction contrast. The two different variants can merge and form a single crystal across adjacent facets as illustrated in **b**, **c** and **d**, and form low energy coherent twins along the direction of the NW as illustrated in panel **f**. In **d** (along the NW axis) and **f** (perpendicular to the NW axis), the different spheres symbolize: As (black), In (white) and Al (grey). Panel **e** shows a case where an incoherent grain boundary appears around the circumference.

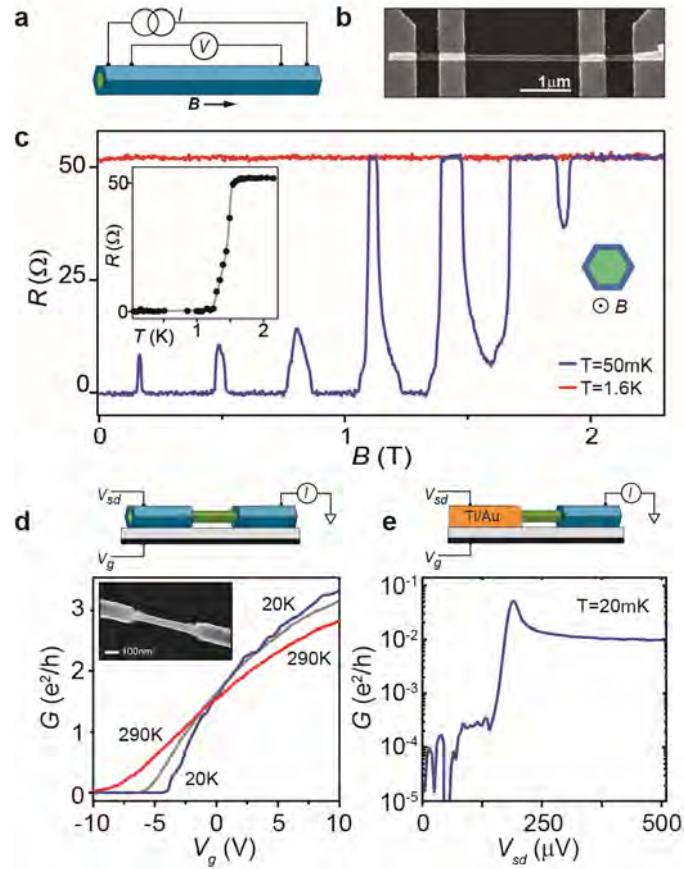

**Fig. 5. Electrical properties of the Al/InAs epitaxial hybrids. a** Schematic illustration of the four-terminal device made from a full-shell Al/InAs nanowire with 13nm thick Al. **b** SEM image of the device. The scale bar is 1µm. **c** Measurements of the four-terminal resistance as a function of magnetic field. The device is superconducting at low fields with Little-Parks peaks appearing at half multiples of flux quanta threading the wire. Inset shows the resistance as a function of temperature with a superconducting transition at ~1.4 K. **d** Conductance as a function of gate-voltage for a device where the InAs core has been exposed. Measurements are shows for various temperatures to investigate the contact barriers between the core and the shell. Upper schematic illustrates the device, and the lower inset shows an SEM micrograph of the central part of the actual device. **e** Top schematic illustrates the device used to perform tunnel spectroscopy of the density of states of the proximitized nanowire shown in the lower panel. The low-bias (sub gap) differential conductance is reduced by two orders of magnitude compared to the above gap (normal state) value.

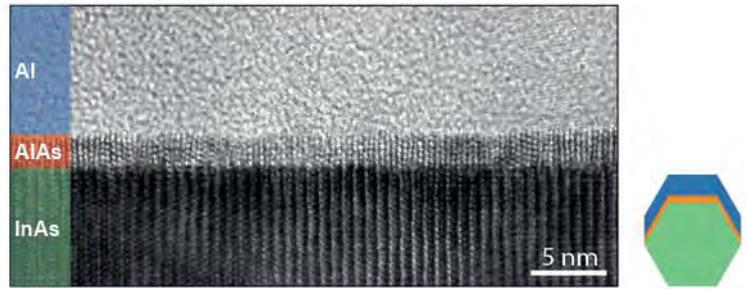

**Fig. 6. Growth of SE/SU interface barriers.**

Transmission electron micrograph of a hybrid structure grown with a ~3nm AlAs high band-gap semiconductor layer separating the InAs core from the Al. The rightmost icon schematically illustrates the structure.

**References:**

[1] Lutchyn, R.M., Sau, J.D. & Das Sarma, S. Majorana Fermions and a Topological Phase Transition in Semiconductor-Superconductor Heterostructures. *Phys. Rev. Lett.* **105**, 077001 (2010).

[2] Oreg, Y., Refael, G. & von Oppen, F. Helical Liquids and Majorana Bound States in Quantum Wires. *Phys. Rev. Lett.* **105**, 177002 (2010).

[3] Mourik, V., Zuo, K., Frolov, S. M., Plissard, S. R., Bakkers, E. P. A. M. & Kouwenhoven, L. P. Signatures of Majorana Fermions in Hybrid Superconductor-Semiconductor Nanowire Devices. *Science* **336**, 1003-1007 (2012).

[4] Das, A., Ronen, Y., Most, Y., Oreg, Y., Heiblum, M., & Shtrikman, H. Zero-bias peaks and splitting in an Al–InAs nanowire topological superconductor as a signature of Majorana fermions. *Nat. Phys.* **8**, 887-895 (2012).

[5] Deng, M. T., Yu, C. L., Huang, G. Y., Larsson, M., Caroff, P. & Xu, H. Q. Anomalous Zero-Bias Conductance Peak in a Nb–InSb Nanowire–Nb Hybrid Device. *Nano Lett.* **12**, 6414-6419 (2012).

[6] Churchill, H. O. H., Fatemi, V., Grove-Rasmussen, K., Deng, M. T., Caroff, P., Xu, H. Q. & Marcus, C. M. Superconductor-nanowire devices from tunneling to the multichannel regime: Zero-bias oscillations and magnetoconductance crossover. *Phys. Rev. B* **87**, 241401 (2013).

[7] Stanescu, T.D. & Das Sarma, S. Superconducting proximity effect in semiconductor nanowires. *Phys. Rev. B* **87**, 180504 (2013).

[8] Takei, S., Fregoso, B. M., Hui, H.-Y., Lobos, A. M. & Das Sarma, S. Soft Superconducting Gap in Semiconductor Majorana Nanowires. *Phys. Rev. Lett.* **110**, 186803 (2013).

[9] Kroemer, H. Quasielectric fields and band offsets: teaching electrons new tricks. *Rev. Mod. Phys.* **73**, 783 (2001).

[10] Lauhon, L. J., Gudiksen, M. S., Wang, D., & Lieber, C. M., Epitaxial core-shell and core-multishell nanowire heterostructures. *Nature* **420**, 57-61 (2002).

[11] Björk, M. T. *et al*. One-dimensional steeplechase for electrons realized, *Nano Lett.* **2**, 87-89 (2002).

[12] Heiss, M. et al., Self-assembled quantum dots in a nanowire system for quantum photonics. *Nat. Mater.* **12**, 439-444 (2013).

[13] Wallentin, J. *et al.*, InP Nanowire Array Solar Cells Achieving 13.8% Efficiency by Exceeding the Ray Optics Limit. *Science* **339**, 1057-1060 (2013).

[14] Borg, M. *et al*., Vertical III-V nanowire device integration on Si(100). *Nano Lett.* **14**, 1914-1920 (2014).

[15] Li, M. *et al*., Bottom-up Assembly of Large-area Nanowire Resonator Arrays. *Nat. Nanotechnol.* **3**, 88-92 (2008).

[16] Pilkington, S. J. & Missous, M., The growth of epitaxial aluminum on As containing compound semiconductors. *J. Cryst. Growth* **196**, 1-12 (1999)

[17] Dick, A. K. *et al*. Synthesis of branched 'nanotrees' by controlled seeding of multiple branching events. *Nat. Mater.* **3**, 380-384 (2004).

[18] Zheleva, T., Jagannadham, K. & Narayan, J. Epitaxial growth in large-lattice mismatch systems. *J. Appl. Phys.* **75**, 860 (1994).

[19] Krogstrup, P. *et al.*, Advances in the theory of nanowire growth dynamics, J. Phys. D: Appl. Phys. **46** 313001 (2013)

[20] Little, W. A. & Parks, R. D. Observation of Quantum Periodicity in the Transition Temperature of a Superconducting Cylinder. *Phys. Rev. Lett.* **9**, 9 (1962).

[21] Liu, Y., Zadorozhny, Y., Rosario, M.M., Rock, B.Y., Carrigan, P.T. & Wang, H. Destruction of the Global Phase Coherence in Ultrathin, Doubly Connected Superconducting Cylinders. *Science* **294**, 2332-2334 (2001).

[22] Staley, N.E. & Liu, Y. Manipulating superconducting fluctuations by the Little-Parks-de Gennes effect in ultrasmall Al loops. *PNAS* **11**, 14819-14823 (2012).

[23] Meservey, R. & Tedrow, P. Properties of Very Thin Aluminum Films. *Jour. Appl. Phys.* **42**, 51 (1971).

[24] Chang, W., Albrecht, S.M., Jespersen, T.S., Kuemmeth, F., Krogstrup, P., Nygård, J. & Marcus, C.M. *Nat Nanotechnol.* Accepted (2014).

[25] Rainis, D. & Loss, D. Majorana qubit decoherence by quasiparticle poisoning. *Phys. Rev. B* **85,** 174533 (2012).

[26] Joyez, P., Lafarge, P., Filipe, A., Esteve, D. & Devoret, M.H. Observation of parity-induced suppression of Josephson tunneling in the superconducting single electron transistor. *Phys. Rev. Lett.* **72**, 2458 (1994).

[27] De Gennes, P.G. Boundary Effects in Superconductors. *Rev. Mod. Phys.* **36**, 225 (1964).

[28] Doh, Y.J., van Dam, J.A., Roest, A.L., Bakkers, E. P. A. M., Kouwenhoven, L.P. & De Franceschi, S. Tunable Supercurrent Through Semiconductor Nanowires. *Science* **309**, 272-275 (2005).

[29] van Dam, J.A., Nazarov, Y.V., Bakkers, E. P. A. M., De Franceschi, S. & Kouwenhoven, L.P. Supercurrent reversal in quantum dots. *Nature* **442**, 667-670 (2006).

[30] Momma, K & Izumi, F, "VESTA 3 for three-dimensional visualization of crystal, volumetric and morphology data," J. Appl. Crystallogr., 44, 1272 (2011)

**Acknowledgement**

We wish to thank B. Wenzell, L. Schulte and J. B. Wagner for TEM sample preparation, W. Zhang for assistance on TEM and EDX analyses, G. Ungaretti for substrate preparation and C.B. Sørensen for technical assistance. We acknowledge financial support by Microsoft Project Q, EU FP7 project SE2ND (no. 271554), the Danish Strategic Research Council, the Danish Advanced Technology Foundation, the Carlsberg Foundation and the Lundbeck Foundation. The Center for Quantum Devices is supported by the Danish National Research Foundation.

**Supplementary information for 'Semiconductor-Superconductor Nanowire Epitaxy'**

P. Krogstrup[1,*], N. L. B. Ziino[1], W. Chang[1], S. M. Albrecht[1], M. H. Madsen[1], E. Johnson[1,2], J. Nygård[1,3], C. M. Marcus[1] and T. S. Jespersen[1,*]

[1] Center for Quantum Devices, Niels Bohr Institute, University of Copenhagen, Denmark
[2] Department of Wind Energy, Technical University of Denmark, Risø Campus, Roskilde, Denmark
[3] Nano-science Center, Niels Bohr Institute, University of Copenhagen, Denmark

\* Corresponding Authors: krogstrup@nbi.dk, tsand@nbi.dk

Content

# S1) Morphological evolution during epitaxial growth of a NW metal shell

Bottom-up grown nanowires (NWs) are far from being equilibrium shape crystals and are a result of far from equilibrium growth kinetics. Also the growth of a uniform metal shell on the facets of such NWs in general require far from equilibrium conditions, where in particular, the surface kinetics of adatoms plays a key role in the evolution of the crystal morphology. In this section, we discuss the role of adatom diffusion on the overall morphology of the metal shell formation. In Fig.S1 we give an overview of the basic stages during the epitaxial grain growth evolution of a thin metal phase.

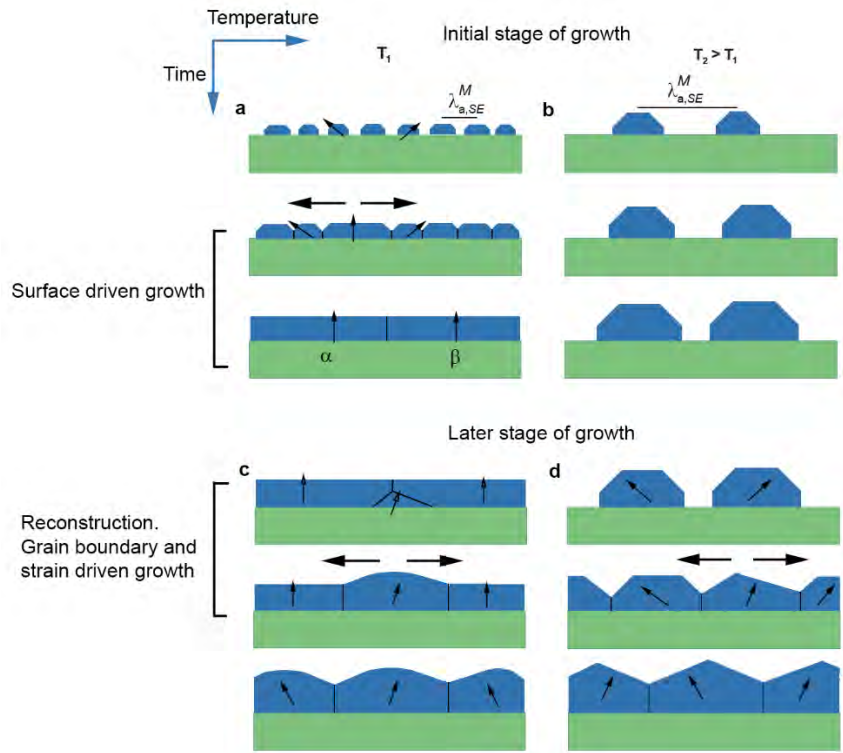

**Fig. S1 Main stages during epitaxial grain growth. a,b** An illustration of the initial stages of the metal growth evolution at relatively low and high temperatures, respectively. In **a**, a low substrate temperature $T_1$ gives small and closely spaced metal grains due to a relatively low adatom mobility, described by an average adatom diffusion length $\lambda_{a,M}^{SE}$ of species $M$ on a facet of $SE$. The small grains will merge into a thin film, where if the film is thin enough, the grains with the lowest surface energy will grow on expense of the grains with higher surface energy (as indicated by the arrows). In **b**, $\lambda_{a,M}^{SE}$ is larger due to a higher temperature, $T_2$, which give larger grains separated further apart. **c**, **d** Continuing growth, both temperature regimes may evolve into new preferred crystal orientations as the role of the grain boundaries and strain contributions become increasingly important with increasing film thickness.

A low substrate temperature promotes formation of small and closely spaced islands because the adatom mobility at the *SE* surface is small (Fig. S1 a). As the islands grow bigger, they will merge into a thin film, and if the film is thin enough when the islands merge, minimization of surface energy dictates the out-of-plane grain orientation[1]. For most FFC metals, this will lead to flat and uniform (111) low energy surfaces (as shown for InAs/Al in Fig.S5 and Fig. S6). A higher temperature will increase the adatom mobility, which results in islands spaced further apart, as illustrated in Fig S1 b. When larger islands merge, the film may have exceeded a certain thickness where the role of primarily incoherent grain boundaries and strain fields provides the dominating driving forces for grain growth, and the growth may never pass through the surface driven stage. At the later stages in growth (as illustrated in Fig. S2 c, d), for both relative low and high temperatures, there can be nucleation of new preferential orientations and reconstruction of the metal phase.

Thus, the overall morphological evolution during growth depends on the adatom mobility, and it is therefore relevant to derive an expression for the mean adatom diffusion length, $\lambda_a$, as function of the basic control parameters, the substrate temperature $T$ and beam flux $f$. We will assume random walk, which implies that the mean distance an adatom travels on a surface $j$, between a 'birth' position (where the atom hits the surface) and a 'death' position (where the adatom is incorporated in the crystal) can be written as

$$\lambda_{a,j} \propto \sqrt{D_{a,j} \tau_{a,j}} \qquad (1)$$

Here $D_a$ and $\tau_a$ are the mean adatom diffusivity and adatom lifetime, respectively. The values of $D_a$ and $\tau_a$ will change significantly from the initial stage of growth where the adatoms 'feel' the surface of *SE*, to later stages where the influence from the *SE* is negligible. However, in both cases, using the theoretical approach in ref.[2], the diffusion length of an adatom state of single species with no probability of being desorbed from the surface, can be written as; $\lambda_a \propto c_{inc}^{-\frac{1}{2}} \exp\left(-\frac{\delta h_{aa} - \delta h_{as}}{2 k_B T}\right)$. Here $c_{inc}$ is the density of incorporation sites (defined as being thermodynamically stable sites), $\delta h_{aa}$ is the transistion state enhalpy barrier between two adjacent adatom sites, and $\delta h_{as}$ is the transistion state enhalpy barrier for incorporation of an adatom at an incorporation site. The density of incorporation sites can be written as $c_{inc} \propto \exp\left(\dfrac{\delta\mu_a - \delta\mu_M}{k_B T}\right)$, where $\delta\mu_a \propto k_B T \ln(\rho_a)$ is a measure of the chemical potential of an isomorphic and ideal adatom state with adatom concentration, $\rho_a$, and $\delta\mu_M$ is the chemical potential of the incorporation site. Assuming that the incorporation barrier is negligible, the mean distance an adatom travels on a surface before it is incorporated writes

$$\lambda_a(\rho_a, T) \propto \rho_a^{-\frac{1}{2}} \exp\left(-\dfrac{\delta h_{aa} - \delta\mu_M}{2 k_B T}\right) \qquad (2)$$

It is obvious from equation (2), that $T$ and $\delta\mu_M$ play a crucial role due to the exponential dependence, while the influence of the beam flux, $f$, is less obvious and can in principle only be solved using numerical simulations through its dependence on $\rho_a$.[2] But the general trend is, increasing $f$ will increase $\rho_a$, which means that $\lambda_a$ is decreasing according to equation (2), which again will lower $\rho_a$ due to a faster incorporation rate. Thus, $f$ is at the most proportional to $\rho_a$ and therefore not as crucial a parameter in the sense of controlling the morphology as the temperature. However, according to this derivation a higher beam flux and a lower temperature decrease the adatom diffusion length. In Fig. S2 S2 we show a top view image of kinked <1-100> InAs NWs with a nominal 20 nm thick Al layer. The Al which looks like pearls on a string are formed due to a short substrate temperature increase from -30°C to approximately 90°C and back to room temperature (as measured on the thermocouple). In accordance with the above discussion, these crystals form due to an increased mobility of the adatoms at higher temperature that makes them able to form shapes closer to equilibrium.

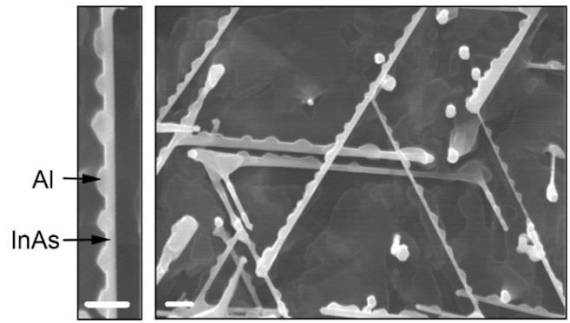

**Fig. S2. Al dewetting.** Top view SEM images of <1-100> InAs NWs grown horizontally on a (111)B InAs substrate, with Al nanocrystals almost equally separated along the NW length. Left, a zoom on a single NW, which demonstrates the Al dewetting on the InAs at elevated temperatures. Scalebars are 300 nm.

## S2) Grain growth kinetics of a metal shell

The metal phase formation consists of both vapor-solid growth and grain growth taking place in parallel, as illustrated in Fig. S1. In this section, we consider the grain growth kinetics once the metal phase is fully covering the semiconductor NW facet or facets, and develop a general continuum model based on thermodynamical parameters, which relates directly to the parameters controlled by the crystal grower.

When desorption is negligible, as in the case of low temperature growth of metals, the vapor-solid transition is in fact a vapor-adatom transition followed by a series of adatom-solid and solid-adatom transitions, under which minimization of surface energy drives the crystal shape evolution[3]. Likewise, minimization of grain boundary energy drives the grain growth[4] and crystal reconstructions, where the transitions take place across the moving grain boundaries. As we will see in the discussion below, it is helpful to construct an applicable theoretical framework to get a more complete picture of the mechanisms driving the growth. We are specifically interested in finding an expression for the growth rate of a given grain as a function of its orientation and morphology.

To describe the grain growth kinetics, we follow the general formalism for material kinetics proposed in ref.[2], which state that flux of atoms through an unit area of grain boundary from a local (continuum) state $p$ to a local state $q$, is given by:

$$\Gamma_{pq} \propto \begin{cases} c_p \exp\left(-\dfrac{\delta g_{pq} - \delta \mu_p}{k_B T}\right) & \text{if } \delta g_{pq} \geq \delta \mu_p \\ c_p & \text{if } \delta g_{pq} \leq \delta \mu_p \end{cases} \quad (3)$$

Here, $c_p$ is the concentration and can therefore be either a constant or zero for single component grain growth depending on whether or not the grain $M'$ occupies the space at $p$. The transition state barrier, $\delta g_{pq}$, is the maximum free energy increase when bringing an 'average atom' from the state $p$ to the state $q$ (Note that the transition rate from p to q is independent of the state of q). $\delta \mu_p$ is the local out of equilibrium chemical potential of $p$ relative to a chosen equilibrium reference state for the system. Assuming detailed balance at equilibrium reference conditions and barrier limited kinetics ($\delta g_{pq} \geq \delta \mu_p$), the effective flux (forward minus backward flux) from $p$ to $q$ is given as, $\Delta\Gamma_{pq} \propto \exp\left(-\frac{\delta g_{pq}}{k_B T}\right)\left(\exp\left(\frac{\delta\mu_q}{k_B T}\right) - \exp\left(\frac{\delta\mu_p}{k_B T}\right)\right)$. Now, we will here not distinguish between local states within a grain, but rather establish an expression for an average growth rate between two grains $M'$ and $\bar{M}$, where we put $\bar{M}$ as an average reference state of the total metallic phase. With this, the rate is

$$\Delta\Gamma_{M'\bar{M}} \propto \exp\left(-\frac{\delta g_{M'\bar{M}}}{k_B T}\right)\left(\exp\left(\frac{\delta\mu_{M'}}{k_B T}\right) - 1\right) \tag{4}$$

where the growth of a given grain is solely driven by the difference in chemical potentials, $\delta\mu_{M'} = \frac{\partial G}{\partial n_{M'}} - \mu_{\bar{M}}$, across the boundaries to the neighbouring grains, with $G$ being the total Gibbs free energy, $n_{M'}$ the number of atoms in $M'$ and $\mu_{\bar{M}}$ the average chemical potential of the NW metal phase. When $\delta\mu_{M'} = 0$, the grain size of $M'$ stays constant on average, according to equation (4), while $\delta\mu_{M'} > 0$ and $\delta\mu_{M'} < 0$ imply an average driving force for elimination and expansion, respectively. This sets the basic framework for analyzing growth mechanisms of the grains. Left is now to find expressions for the thermodynamic driving force $\delta\mu_{M'}$ that governs the growth evolution.

For single component materials, it is sufficient to find an expression for changes in the excess Gibbs free energy upon a transition between $M'$ and $\bar{M}$:

$$\delta\mu_{M'}^X = \frac{\partial G_{ex}^X}{\partial X}\frac{\partial X}{\partial n_{M'}} - \mu_{\bar{M}}^X \tag{5}$$

where $X$ is a parameter describing a corresponding change in the crystal shape. To explain the mechanisms behind the growth evolution, it is convenient to split the free energy minimization into two main parts as:

$$\frac{\partial G_{ex}^X}{\partial n_{M'}} = \sum_i \gamma_i \frac{\partial A_i}{\partial X}\frac{\partial X}{\partial n_{M'}} + \frac{\partial \mathrm{E}}{\partial X}\frac{\partial X}{\partial n_{M'}} \tag{6}$$

The first sum arise from an excess free energy related to the chemical interaction (bonding) at the interfaces involved, where $\gamma_i$ and $A_i$ are the interface energy density and area of interface $i$, respectively. The second term emerges from changes in the strain energy $\mathrm{E}$ within $M$ and $SE$ due to changes in $n_{M'}$. For a full description of the growth kinetics, the change in excess free energy needs to be described in terms of a complete set of independent parameters, $\{X\}$, which fully describe the shape of the crystal. However, we can capture the most important growth mechanisms by examining the growth of a grain $M'$ on a single facet (and therefore relevant for planar growth as well) using only two parameters: $\{X\} \in \{h_{M'}, R_{M'}\}$, where $h_{M'}$ is the average grain thickness and $R_{M'}$ is the average in-plane radius of curvature, see Fig. S3 a, b. Depending on the sign of $\delta\mu_{M'}^{h_{M'}}$, material will move from $M'$ to $\bar{M}$ via adatom surface diffusion, and depending on the sign of $\delta\mu_{M'}^{R_{M'}}$, material will move from $M'$ to $\bar{M}$ across the grain boundaries.

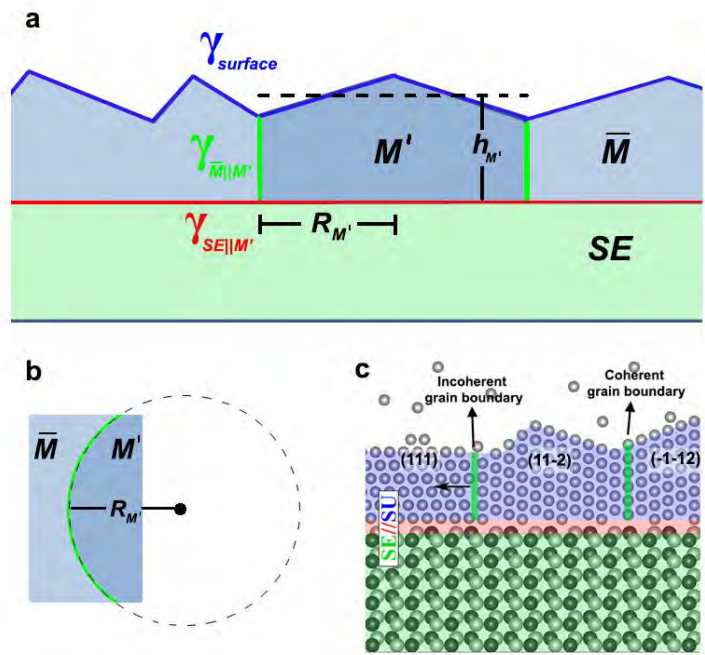

**Fig. S3 Thin film growth of a single facet. a** A view along an in-plane direction of the metal (blue) - semiconductor (green) material. The average chemical potential of a given grain $M'$ is described in terms of the excess free energies associated with the three interfaces involved, the metal surface, the grain boundaries ($\bar{M} \| M'$) and the semiconductor-metal interface ($SE \| M'$). $h_{M'}$ is as indicated the average thickness of the grain. **b** An out of plane illustration of a grain boundary, where the grain boundary driven growth depends on the local curvature, but here used as an average curvature $R_{M'}$. **c** An illustration of typical coherent and in-coherent grain boundaries seen in the NWs.

In the thin film limit, the coherent strain energy associated with the domain mismatch at the $SE \| M'$ interface can be written as[5]:

$$E \simeq \frac{\mathbf{S}}{1-\nu}\left(\varepsilon_{\parallel}^2 + \varepsilon_{\perp}^2 + 2\nu\varepsilon_{\parallel}\varepsilon_{\perp}\right)h_{M'}\pi R_{M'}^2, \qquad (7)$$

where $\mathbf{S}$ is the shear modulus, $\nu$ is Poisson's ratio and $\varepsilon_i$ is the strain induced from the lattice or domain mismatch in the i'th interfacial direction. The chosen parameters need to be related to the number of atoms in the grain, which in the single facet case is simply given as: $n_{M'} \propto \frac{h_{M'} R_{M'}^2}{\Omega_M}$, where $\Omega_M$ is the atomic volume. This means that the second term in equation (6) is set by,

$\frac{\partial E}{\partial h_{M'}} \frac{\partial h_{M'}}{\partial n_{M'}} = \frac{\mathbf{S}\varepsilon_{M'}^2 \Omega_M}{(1-\nu)}$, with strain parameter $\varepsilon_{M'}$ defined as, $\varepsilon_{M'}^2 \equiv \varepsilon_{M',\parallel}^2 + \varepsilon_{M',\perp}^2 + 2\nu\varepsilon_{M',\parallel}\varepsilon_{M',\perp}$.

Hence, the driving force for changes in $h_{M'}$, according to equation (3) and (6), becomes

$$\delta\mu_{M'}^{h_{M'}} = \Omega_M \frac{\gamma_{\bar{M}\parallel M'}^j}{R_{M'}} + \Omega_M \frac{\mathbf{S}\varepsilon_{M'}^2}{(1-\nu)} - \mu_{\bar{M}}^{h_{\bar{M}}} \qquad (8)$$

Equation (8) tell us that if the grain boundary energy associated with $M'$ is higher or the grain is smaller than on average, there is a driving force for moving material from $M'$ to adjacent grains. Also, as expected in the coherent strained regime, a highly strained $M'$ will also contribute to the driving force for making $h_{M'}$ smaller, but this contribution does not depend on the size of $M'$. Changes in the surface morphology may induce changes on the driving force with respect to grain growth driven by changes in $R_{M'}$. Deriving the thermodynamic driving force for changes in $R_{M'}$ for single facet growth gives:

$$\delta\mu_{M'}^{R_{M'}} = \Omega_M \frac{\gamma_{surface,M'} + \gamma_{SE\parallel M'}}{h_{M'}} + \Omega_M \frac{\gamma_{\bar{M}\parallel M'}^j}{R_{M'}} + \frac{\Omega_M \mathbf{S}\varepsilon^2}{(1-\nu)} - \mu_{\bar{M}}^{R_{\bar{M}}} \qquad (9)$$

Equation (9) serves as a platform for qualitative analysis of the grain growth and the preferred crystal orientations at different stages during growth. It tells us that at in the beginning of growth, the first term dominates (when $h_{M'}$ is small) and the grains orientate according to the lowest surface energies for the out of plane orientation and according to the lowest $SE\|M'$ interface energy for the in-plane orientation[4]. As the metallic phase grows thicker, the grain boundary term (second term) and the strain term (the third term) start to play an important role, which may change the preferred grain orientations if $\delta\mu_{M'}^{R_{M'}}$ is large enough to overcome the barrier for reconstruction. In general, incoherent grain boundaries induce a high driving force for grain boundary elimination according to the second term in equation (9). Thus if for example a (111) and a (11-2) oriented grain form a boundary which is incoherent, as sketch in Fig. S3c, and there will be a high driving force for growth of the grain with the lowest chemical potential. A coherent twin boundary such as between two {11-2} oriented grains has a low excess energy density and they are locked kinetically much easier than the incoherent boundaries.

The above derivation of the driving forces applies for single facet growth or for more facets as long as $h_M \ll d_{NW}$, where $d_{NW}$ is the semiconductor NW diameter. For thicker shells with more facets, especially full shell growth, we need to take account of the fact that the shell is growing roughly as a cylinder. To describe the evolution of this type of growth, we need more shape parameters, including $d_{NW}$, however, as the essential mechanisms is captured by the above discussion, more complicated and detailed derivations is out of the scope of this study.

# S3) Epitaxial SE/M interfaces - interfacial bicrystal symmetries and degenerate grain orientations

We will here analyse the epitaxial relation between the semiconductor and metal phase. Such phases are in general incommensurate in their relaxed states, but domain matching and strain can cause the two crystals to be in local registry. If we only consider the interfacial structure, there will be preferred relative orientations, which only depend on the relative lattice constants and plane rotational symmetries of the constituents (See for example Novaco & Mctague's[6] discussion on single plane epitaxial ordering). However, as discussed in section S2, it is not only the epitaxial ordering that determines the relative orientations of composite crystals but also excess energies associated with surfaces and grain boundaries. Based on TEM inspection of many different types of InAs/Al NWs, it is clear that there exists a preferred out-of-plane orientation of the Al phase, for a given Al thickness and type of core-shell hybrid structure. Moreover, for each out-of-plane orientation there seems to be a very limited number of in-plane orientations present. Grey & Bohr[7] formulated a principle of epitaxial rotation, which gives all the equivalent interfaces of unstrained structures as a function of relative rotation. However, our results suggest (as discussed in the main text) that the interfaces try to minimize the energy by forming small low energy domains on the cost of straining their structures. Thus, as we discussed the overall mechanisms leading to certain out of plane orientations in the previous sections (S1 and S2), we will here suggest a simple way to analyse the in-plane epitaxial matching in terms of a given out-of-plane orientation. For a general theoretical framework for bicrystal symmetries based on the bulk symmetry groups of the component lattices we refer to Pond & Bollmann[8]. We will here examine the bilayer rotational symmetries normal to the interface of two joining arbitrary crystals, *SE* and *M*, where we consider *SE* as a fixed reference. For a given out of plane orientation of *M*, there exists a given number of degenerate crystal orientations, i.e. with indistinguishable interfacial planes, but with distinguishable crystal orientations in *M*. We will call the crystal orientations that correspond to a given type of *SE/M* interface for the *variants* of *M*, and describe the epitaxial relation and ordering of the *SE/M* interfaces in terms of domain matching. That is, when a lattice mismatch of a given *SE/M* material combination is large, interfacial domains - consisting of $n_M$ and $n_{SE}$ interfacial units of *M* and *SE* - form, in order to reduce the stress associated with the mismatch[5]. For a given out of plane orientation of *M*, there exists a set of variants with a certain in-plane orientation that minimizes the free energy. From the set of variants, the low energy grain boundaries can be obtained (see ref.[9] for details on grain boundary orientations corresponding energies). Based on these statements, we will provide a simple general framework to describe the in-plane orientation of *M*, in terms of symmetries of *M* and *SE*. For NWs with rough surfaces, the orientation of the grains becomes more random, see Fig. S6 c. We are here only interested in the planar *SE* surfaces, and we neglect for simplicity what we assume to be small energy differences arising from polarity in the *SE*. The order of the plane rotational symmetries (*PRS*) of *SE* and *M* along a given crystal axis $i$ will be denoted, $C_{i,SE}^{PRS,\perp}$ and $C_{i,M}^{PRS,\perp}$, respectively. The superscript $\perp$ specifies that we are considering the symmetries in the transverse direction, i.e. a single interface. Note that the order of *PRS*'s are not necessarily the same as those of the corresponding symmetry operations $C_{i,SE}^{\perp}$ and $C_{i,M}^{\perp}$ of the bulk crystallographic point groups, because an atomic plane can have higher symmetry than the corresponding bulk operation. Thus, for a single *SE/M* interface $i$, the number of distinguishable degenerate crystal orientations in *M* are given as

$$m_i^{\perp} = \frac{\mathbb{N}\left(C_{i,SE}^{PRS,\perp}, C_{i,M}^{PRS,\perp}\right)}{C_{i,M}^{\perp}} \quad (10)$$

where $\mathbb{N}\left(C_{i,SE}^{PRS}, C_{i,M}^{PRS}\right)$ is the least common multiple of $C_{i,SE}^{PRS}$ and $C_{i,M}^{PRS}$. Equation (10) can be visualized using the bicrystal symmetry diagrams presented in Fig. S4.

**Fig. S4 Examples of symmetry diagrams for degenerate single plane bicrystals.** The *SE* plane symmetry is a fixed reference, while the *M* plane and bulk symmetry rotates around a point normal to the interface. If a given rotation gives the same plane symmetry but different bulk configuration, it is a variant, or specified in another way: the number of different *M* bulk orientations for a given interface pattern defines the number of degenerate variants, in this single plane symmetry limit.

We now consider symmetries along the NW axis, $\parallel$, and restricting the single facets to have only one class of variants. We will also assume that the cross sectional crystal shape of the NW follows the Wulff shape containing only the highest symmetry facets. Then we can say that if $C_{SE,i}^{\parallel} = C_{M,i}^{\parallel}$ there can be $m_i^{\perp} - 1$ different types of grain boundaries across the facets. If these grain boundaries are incoherent, they induces a high driving force for grain growth to eliminate the boundary across the facets, according to equation (7) and as seen in Fig.4a in the main text, where the grain growth does not introduce new classes of variants due to the bicrystal symmetries. More general, if a given variant of the transverse dimension falls into symmetry operations of the parallel dimension, it will not contribute to new class of variants.

## S4) Epitaxial match for other NW material combinations

A very large number of planes and relative orientations are candidates as the preferred structure for given material combinations and phase thicknesses. Also, surface diffusion lengths of 'metal' adatoms on semiconductors, as discussed in section S1, are generally not available in the literature. Thus, it is very difficult to make general predictions for material combinations and growth conditions that will produce uniform semiconductor-metal NW heterostructures with a good epitaxial match. However, in the thick shell limit, where strain and grain boundary driven growth dominates, the lowest energy configuration is likely to be when the crystal symmetry of the metal follows the overall cross sectional shape of the NW core. That is, if the NW core has a hexagonal (6-fold) cross section, the metal phase will orientate, if possible, with a 6-fold symmetry along the NW length, in order to keep coherency across the adjacent facets. Thus, if the SE and M crystals have similar symmetry groups (ZB or WZ and FCC) they orientate along the same type of symmetry classes, especially if the domain mismatch in these directions is not too large.

As discussed in the main text, for nanowires grown in the conventional $[0001]_{WZ}$ / $[111]_{ZB}$ direction with $\{1\text{-}100\}/\{11\text{-}2\}$ type faceting, a cubic metal phase with the $<11\text{-}2>$ out-of-plane orientation and $[111]$ along the nanowire axis is unique, in that its overall symmetry allows large single crystal segments with simultaneous epitaxial match on all facets of the nanowire. Thus, it is natural to expect, that if this orientation matches the semiconductor for a particular metal, it is likely to form in the thick film limit. Table 1, 2, and 3 list the domain strains $\varepsilon_\parallel = \varepsilon_\perp$ for a range of metals grown on the important cases of InAs, InSb, and GaAs. In the general notation $\left(\dfrac{n_{M,\parallel}}{n_{SE,\parallel}}, \varepsilon_\parallel\right) \times \left(\dfrac{n_{M,\perp}}{n_{SE,\perp}}, \varepsilon_\perp\right)$, we distinguish between interfacial match of interfacial units and the corresponding strain along the length and along the transverse direction to the NW, as expected from relaxed bulk values. However, if the ZB and FCC orientate along the same type symmetry classes, the two numbers are identical in the parallel and perpendicular directions. We emphasize that the tables below are only suggestions for possible feasible material combinations – and as discussed in the main text and section S1 and S2 many other factors can play important roles in the formation of a particular crystal orientation, and combinations without match in the tables may form epitaxial interfaces in other orientations.

## S4.1) InAs/ metal bicrystal match

| Domain fraction ZB//FCC | Lattice const. | ½ | 1/3 | 2/3 | 1/4 | 3/4 | 1/5 | 2/5 | 3/5 | 4/5 |
|---|---|---|---|---|---|---|---|---|---|---|
| FCC metal | | | | | | | | | | |
| Al | 4.05 | 25.2 | 50.1 | 0.3 | 62.6 | 12.2 | 70.1 | 40.2 | 10.2 | 19.7 |
| Ca | 5.58 | 45.7 | 63.8 | 27.6 | 72.9 | 18.6 | 78.3 | 56.6 | 34.9 | 13.1 |
| Ni | 3.52 | 13.9 | 42.6 | 14.7 | 57.0 | 29.1 | 65.6 | 31.2 | 3.3 | 37.7 |
| Cu | 3.61 | 16.1 | 44.1 | 11.9 | 58.0 | 25.9 | 66.4 | 32.9 | 0.7 | 34.3 |
| Sr | 6.08 | 50.2 | 66.8 | 33.6 | 75.1 | 25.3 | 80.1 | 60.1 | 40.2 | 20.3 |
| Rh | 3.8 | 20.3 | 46.9 | 6.3 | 60.1 | 19.6 | 68.1 | 36.2 | 4.3 | 27.5 |
| Pd | 3.89 | 22.1 | 48.1 | 3.8 | 61.1 | 16.8 | 68.9 | 37.7 | 6.6 | 24.6 |
| Ag | 4.09 | 25.9 | 50.6 | 1.3 | 63.0 | 11.1 | 70.4 | 40.8 | 11.1 | 18.5 |
| Ce | 5.16 | 41.3 | 60.9 | 21.7 | 70.6 | 11.9 | 76.5 | 53.0 | 29.6 | 6.1 |
| Yb | 5.49 | 44.8 | 63.2 | 26.4 | 72.4 | 17.2 | 77.9 | 55.9 | 33.8 | 11.7 |
| Ir | 3.84 | 21.1 | 47.4 | 5.2 | 60.6 | 18.3 | 68.4 | 36.9 | 5.3 | 26.2 |
| Pt | 3.92 | 22.7 | 48.5 | 3.0 | 61.4 | 15.9 | 69.1 | 38.2 | 7.3 | 23.6 |
| Au | 4.08 | 25.8 | 50.5 | 1.0 | 62.9 | 11.4 | 70.3 | 40.6 | 10.9 | 18.8 |
| Pb | 4.95 | 38.8 | 59.2 | 18.4 | 69.4 | 8.2 | 75.5 | 51.0 | 26.6 | 2.1 |
| Ac | 5.31 | 43.0 | 62.0 | 23.9 | 71.5 | 14.4 | 77.2 | 54.4 | 31.5 | 8.7 |
| Th | 5.08 | 40.4 | 60.2 | 20.5 | 70.2 | 10.6 | 76.1 | 52.3 | 28.4 | 4.6 |
| ZB//BCC | | | | | | | | | | |
| Li | 3.49 | 13.2 | 42.1 | 15.7 | 56.6 | 30.2 | 65.3 | 30.6 | 4.2 | 38.9 |
| Na | 4.23 | 28.4 | 52.3 | 4.5 | 64.2 | 7.4 | 71.4 | 42.7 | 14.1 | 14.6 |
| K | 5.23 | 42.1 | 61.4 | 22.8 | 71.0 | 13.1 | 76.8 | 53.7 | 30.5 | 7.3 |
| V | 3.02 | 0.3 | 33.1 | 33.7 | 49.8 | 50.5 | 59.9 | 19.8 | 20.4 | 60.5 |
| Cr | 2.88 | 5.2 | 29.9 | 40.2 | 47.4 | 57.8 | 57.9 | 15.9 | 26.2 | 68.3 |
| Fe | 2.87 | 5.5 | 29.6 | 40.7 | 47.2 | 58.3 | 57.8 | 15.6 | 26.7 | 68.9 |
| Rb | 5.59 | 45.8 | 63.9 | 27.7 | 72.9 | 18.7 | 78.3 | 56.6 | 35.0 | 13.3 |
| Nb | 3.3 | 8.2 | 38.8 | 22.4 | 54.1 | 37.7 | 63.3 | 26.6 | 10.2 | 46.9 |
| Mo | 3.15 | 3.8 | 35.9 | 28.2 | 51.9 | 44.2 | 61.5 | 23.1 | 15.4 | 53.9 |
| Cs | 6.05 | 49.9 | 66.6 | 33.2 | 75.0 | 24.9 | 80.0 | 59.9 | 39.9 | 19.9 |
| Ba | 5.02 | 39.7 | 59.8 | 19.5 | 69.8 | 9.5 | 75.9 | 51.7 | 27.6 | 3.5 |
| Eu | 4.61 | 34.3 | 56.2 | 12.4 | 67.1 | 1.4 | 73.7 | 47.4 | 21.2 | 5.1 |
| Ta | 3.31 | 8.5 | 39.0 | 22.0 | 54.2 | 37.3 | 63.4 | 26.8 | 9.8 | 46.4 |
| W | 3.16 | 4.1 | 36.1 | 27.8 | 52.1 | 43.8 | 61.7 | 23.3 | 15.0 | 53.4 |

**Tabel 1. Domain matches for InAs ZB with different cubic metals, in the case when the component crystals are aligned along the same type of cubic directions. The best matched combinations are highlighted.**

## S4.2) InSb/metal bicrystal match

| Domain fraction ZB/FCC | | ½ | 1/3 | 2/3 | 1/4 | 3/4 | 1/5 | 2/5 | 3/5 | 4/5 |
|---|---|---|---|---|---|---|---|---|---|---|
| FCC metal | lattice const | | | | | | | | | |
| Al | 4.05 | 20.0 | 46.7 | 6.7 | 60.0 | 20.0 | 68.0 | 36.0 | 4.0 | 28.0 |
| Ca | 5.58 | 41.9 | 61.3 | 22.6 | 71.0 | 12.9 | 76.8 | 53.6 | 30.3 | 7.1 |
| Ni | 3.52 | 8.0 | 38.6 | 22.7 | 54.0 | 38.0 | 63.2 | 26.4 | 10.4 | 47.3 |
| Cu | 3.61 | 10.3 | 40.2 | 19.6 | 55.1 | 34.6 | 64.1 | 28.2 | 7.7 | 43.6 |
| Sr | 6.08 | 46.7 | 64.5 | 29.0 | 73.4 | 20.1 | 78.7 | 57.4 | 36.1 | 14.8 |
| Rh | 3.8 | 14.8 | 43.2 | 13.7 | 57.4 | 27.9 | 65.9 | 31.8 | 2.3 | 36.4 |
| Pd | 3.89 | 16.7 | 44.5 | 11.0 | 58.4 | 24.9 | 66.7 | 33.4 | 0.1 | 33.2 |
| Ag | 4.09 | 20.8 | 47.2 | 5.6 | 60.4 | 18.8 | 68.3 | 36.6 | 5.0 | 26.7 |
| Ce | 5.16 | 37.2 | 58.1 | 16.3 | 68.6 | 5.8 | 74.9 | 49.8 | 24.7 | 0.4 |
| Yb | 5.49 | 41.0 | 60.7 | 21.3 | 70.5 | 11.5 | 76.4 | 52.8 | 29.2 | 5.6 |
| Ir | 3.84 | 15.6 | 43.8 | 12.5 | 57.8 | 26.5 | 66.3 | 32.5 | 1.2 | 35.0 |
| Pt | 3.92 | 17.4 | 44.9 | 10.2 | 58.7 | 24.0 | 66.9 | 33.9 | 0.8 | 32.2 |
| Au | 4.08 | 20.6 | 47.1 | 5.9 | 60.3 | 19.1 | 68.2 | 36.5 | 4.7 | 27.0 |
| Pb | 4.95 | 34.6 | 56.4 | 12.7 | 67.3 | 1.8 | 73.8 | 47.6 | 21.5 | 4.7 |
| Ac | 5.31 | 39.0 | 59.3 | 18.7 | 69.5 | 8.5 | 75.6 | 51.2 | 26.8 | 2.4 |
| Th | 5.08 | 36.2 | 57.5 | 15.0 | 68.1 | 4.3 | 74.5 | 49.0 | 23.5 | 2.0 |
| BCC | | | | | | | | | | |
| Li | 3.49 | 7.2 | 38.1 | 23.8 | 53.6 | 39.2 | 62.9 | 25.7 | 11.4 | 48.5 |
| Na | 4.23 | 23.4 | 48.9 | 2.1 | 61.7 | 14.9 | 69.4 | 38.7 | 8.1 | 22.5 |
| K | 5.23 | 38.1 | 58.7 | 17.4 | 69.0 | 7.1 | 75.2 | 50.4 | 25.7 | 0.9 |
| V | 3.02 | 7.3 | 28.5 | 43.0 | 46.4 | 60.9 | 57.1 | 14.2 | 28.7 | 71.6 |
| Cr | 2.88 | 12.5 | 25.0 | 50.0 | 43.8 | 68.7 | 55.0 | 10.0 | 35.0 | 80.0 |
| Fe | 2.87 | 12.9 | 24.8 | 50.5 | 43.6 | 69.3 | 54.9 | 9.7 | 35.4 | 80.6 |
| Rb | 5.59 | 42.0 | 61.4 | 22.7 | 71.0 | 13.1 | 76.8 | 53.6 | 30.5 | 7.3 |
| Nb | 3.3 | 1.8 | 34.6 | 30.9 | 50.9 | 47.3 | 60.7 | 21.5 | 17.8 | 57.1 |
| Mo | 3.15 | 2.8 | 31.4 | 37.1 | 48.6 | 54.3 | 58.9 | 17.7 | 23.4 | 64.5 |
| Cs | 6.05 | 46.5 | 64.3 | 28.6 | 73.2 | 19.7 | 78.6 | 57.2 | 35.7 | 14.3 |
| Ba | 5.02 | 35.5 | 57.0 | 14.0 | 67.7 | 3.2 | 74.2 | 48.4 | 22.6 | 3.3 |
| Eu | 4.61 | 29.7 | 53.2 | 6.3 | 64.9 | 5.4 | 71.9 | 43.8 | 15.7 | 12.4 |
| Ta | 3.31 | 2.1 | 34.8 | 30.5 | 51.1 | 46.8 | 60.9 | 21.7 | 17.4 | 56.6 |
| W | 3.16 | 2.5 | 31.7 | 36.7 | 48.7 | 53.8 | 59.0 | 18.0 | 23.0 | 64.0 |

**Tabel 2 Domain matches for InSb ZB with different cubic metals, in the case when the component crystals are aligned along the same type of cubic directions. The best matched combinations are highlighted.**

## S4.3) GaAs/ metal bicrystal match

| Domain fraction ZB/FCC | | ½ | 1/3 | 2/3 | 1/4 | ¾ | 1/5 | 2/5 | 3/5 | 4/5 |
|---|---|---|---|---|---|---|---|---|---|---|
| fcc metal | lattice const | | | | | | | | | |
| Al | 4.05 | 30.2 | 53.5 | 6.9 | 65.1 | 4.7 | 72.1 | 44.2 | 16.2 | 11.7 |
| Ca | 5.58 | 49.3 | 66.2 | 32.5 | 74.7 | 24.0 | 79.7 | 59.5 | 39.2 | 18.9 |
| Ni | 3.52 | 19.7 | 46.5 | 7.1 | 59.8 | 20.5 | 67.9 | 35.8 | 3.6 | 28.5 |
| Cu | 3.61 | 21.7 | 47.8 | 4.4 | 60.9 | 17.4 | 68.7 | 37.4 | 6.0 | 25.3 |
| Sr | 6.08 | 53.5 | 69.0 | 38.0 | 76.8 | 30.3 | 81.4 | 62.8 | 44.2 | 25.6 |
| Rh | 3.8 | 25.6 | 50.4 | 0.8 | 62.8 | 11.6 | 70.2 | 40.5 | 10.7 | 19.0 |
| Pd | 3.89 | 27.3 | 51.6 | 3.1 | 63.7 | 9.0 | 70.9 | 41.9 | 12.8 | 16.3 |
| Ag | 4.09 | 30.9 | 53.9 | 7.9 | 65.4 | 3.7 | 72.4 | 44.7 | 17.1 | 10.6 |
| Ce | 5.16 | 45.2 | 63.5 | 27.0 | 72.6 | 17.8 | 78.1 | 56.2 | 34.3 | 12.4 |
| Yb | 5.49 | 48.5 | 65.7 | 31.4 | 74.3 | 22.8 | 79.4 | 58.8 | 38.2 | 17.6 |
| Ir | 3.84 | 26.4 | 50.9 | 1.9 | 63.2 | 10.4 | 70.6 | 41.1 | 11.7 | 17.8 |
| Pt | 3.92 | 27.9 | 51.9 | 3.9 | 63.9 | 8.2 | 71.2 | 42.3 | 13.5 | 15.4 |
| Au | 4.08 | 30.7 | 53.8 | 7.6 | 65.4 | 3.9 | 72.3 | 44.6 | 16.9 | 10.8 |
| Pb | 4.95 | 42.9 | 61.9 | 23.9 | 71.4 | 14.3 | 77.2 | 54.3 | 31.5 | 8.6 |
| Ac | 5.31 | 46.8 | 64.5 | 29.0 | 73.4 | 20.2 | 78.7 | 57.4 | 36.1 | 14.8 |
| Th | 5.08 | 44.4 | 62.9 | 25.8 | 72.2 | 16.5 | 77.7 | 55.5 | 33.2 | 11.0 |
| BCC | | | | | | | | | | |
| Li | 3.49 | 19.0 | 46.0 | 8.0 | 59.5 | 21.5 | 67.6 | 35.2 | 2.8 | 29.6 |
| Na | 4.23 | 33.2 | 55.5 | 10.9 | 66.6 | 0.2 | 73.3 | 46.5 | 19.8 | 6.9 |
| K | 5.23 | 46.0 | 64.0 | 27.9 | 73.0 | 18.9 | 78.4 | 56.8 | 35.1 | 13.5 |
| V | 3.02 | 6.4 | 37.6 | 24.8 | 53.2 | 40.4 | 62.6 | 25.1 | 12.3 | 49.8 |
| Cr | 2.88 | 1.9 | 34.6 | 30.9 | 50.9 | 47.2 | 60.7 | 21.5 | 17.8 | 57.0 |
| Fe | 2.87 | 1.5 | 34.3 | 31.3 | 50.8 | 47.7 | 60.6 | 21.2 | 18.2 | 57.6 |
| Rb | 5.59 | 49.4 | 66.3 | 32.6 | 74.7 | 24.2 | 79.8 | 59.5 | 39.3 | 19.1 |
| Nb | 3.3 | 14.3 | 42.9 | 14.2 | 57.2 | 28.5 | 65.7 | 31.5 | 2.8 | 37.0 |
| Mo | 3.15 | 10.3 | 40.2 | 19.6 | 55.1 | 34.6 | 64.1 | 28.2 | 7.7 | 43.6 |
| Cs | 6.05 | 53.3 | 68.9 | 37.7 | 76.6 | 29.9 | 81.3 | 62.6 | 43.9 | 25.2 |
| Ba | 5.02 | 43.7 | 62.5 | 24.9 | 71.8 | 15.5 | 77.5 | 55.0 | 32.4 | 9.9 |
| Eu | 4.61 | 38.7 | 59.1 | 18.2 | 69.3 | 8.0 | 75.5 | 50.9 | 26.4 | 1.9 |
| Ta | 3.31 | 14.6 | 43.1 | 13.9 | 57.3 | 28.1 | 65.8 | 31.7 | 2.5 | 36.6 |
| W | 3.16 | 10.5 | 40.4 | 19.3 | 55.3 | 34.2 | 64.2 | 28.4 | 7.3 | 43.1 |

**Tabel 3. Domain matches for GaAs ZB with different cubic metals, in the case when the component crystals are aligned along the same type of cubic directions. The best matched combinations are highlighted.**

## S5) Bending of <1-100>$_{WZ}$||<11-2>$_{ZB}$ type NWs

An example of 30 nm Al single facet shell on the (111)B topfacet of a kinked <1-100> NW. The InAs NWs first grow in the conventional [0001]B direction vertical on the substrate. After reaching the desired length of the stem, the NW is kinked into one of the {1-100} directions perpendicular to the substrate by either a temperature drop or a short supply of Ga. The growth rate of kinked directions is much higher than the conventional growth. For the Al growth, the substrate is orientated with respect to the Al source such that two of the six orientations will only have Al on the topfacet. For the other four directions, Al will growth on two of the four sidefacets. In Fig. S5 it is seen that the wire is bending away from the Al side indicating a tensile/compressive strain the InAs/Al phases. Note that the conventional NWs with half shell Al tend to bend towards the Al, see main text for discussion.

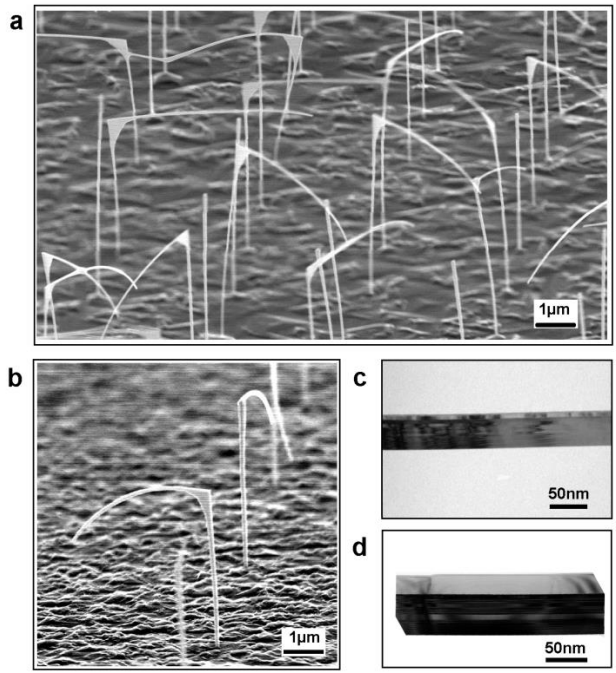

**Fig. S5 Kinked InAs NWs with Al on the topfacet. a,b** Sideview SEM images of (0001)B InAs NW stems (vertical to the substrate) with <1-100> kinked NWs (parallel to the substrate) grown on patterned InAs (111)B substrate. The Al is grown on the (0001)B topfacet of the kinked wires with a thickness of 10nm for the growth shown in **a** and 30nm for the growth shown in **b**, and as shown in **c** and **d** respectively (the same wires as shown in Fig.3 d and e in the main text). This makes the NW bend away from the Al phase due to the small positive domain mismatch $\left(\frac{3_{[11-2]}}{2_{[11-2]_{ZB}}}, 0.3\%\right) \times \left(\frac{3_{[1-10]}}{2_{[1-10]_{ZB}}}, 0.3\%\right)$ whereas a thicker Al appears to bend the wires more strongly.

# S6) Surface driven and grain boundary driven examples of thin two-facet and thick three-facet shells

As mentioned in the main text it is possible to grow the metal phase on selected facets by rotating the facets towards the incoming beam flux. Below, in Figure S6, we show a 'two facet' growth with only ~8-10 nm Al, where the Al phase formation was still in the surface driven growth mode (see discussion on the growth kinetics in section S1). In Fig. S7 we show TEM images of a 'three-facet' growth with approximately 50 nm Al shell, where the final structure forms in the grain boundary driven growth mode, where the WZ and FCC align along the same type of symmetries.

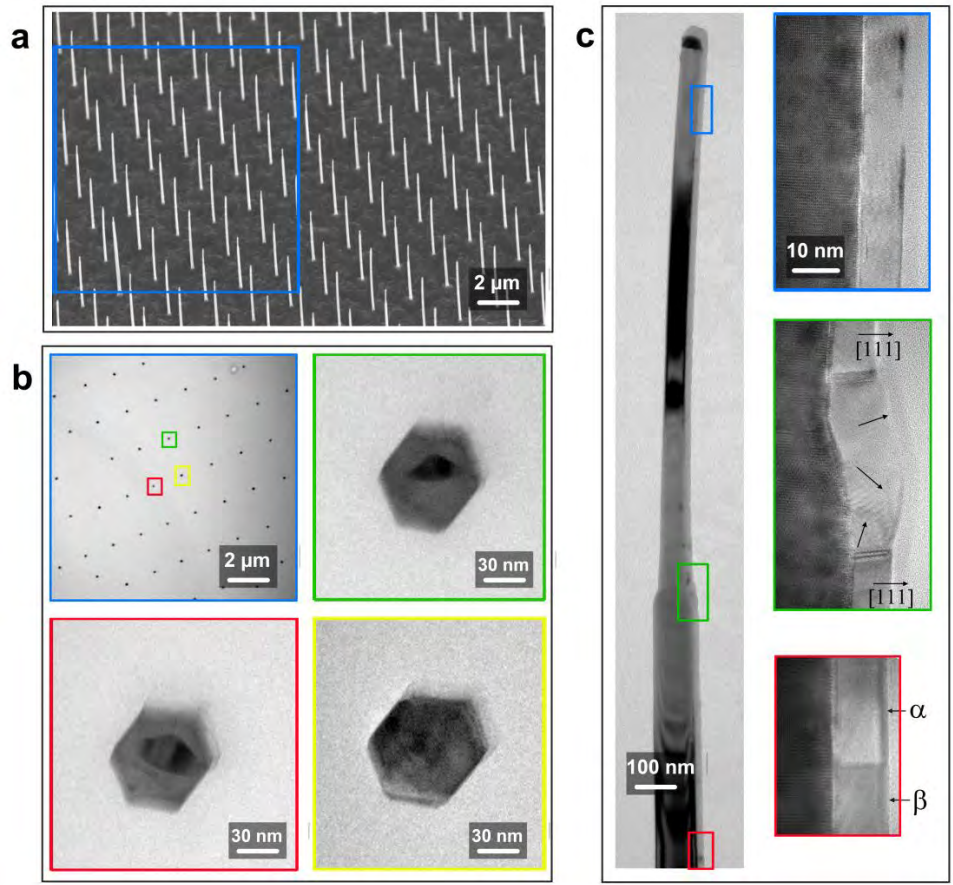

**Fig. S6 Surface driven two-facet thin shell Al/InAs NWs. a** A SEM image of an array of InAs NWs with 10 nm of Al grown on two facets. **b** TEM images of a cross sectional cuts through the NW array in a region similar to the square blue region in **a**. Three examples of individual wires are shown in higher resolution and marked with corresponding squared colour boxes. The Al is covering two facets, which is seen by the light grey contrast at the two top-right facets. **c** TEM images of a single NW from the same growth as in **a** and **b**, where the [111] out of plane orientation all the way along the NW, as seen in blue and red close up boxes, except where the change its diameter (green region). In the red region both variants, $\alpha$ and $\beta$ of the [111] out of plane orientation is shown (these are the only two variants of this type of interface according to equation (10)). In the green region the NW surface is rough and the orientation of the Al phase gets mixed with no clear orientation, indicated by the arrows pointing in various [111] directions.

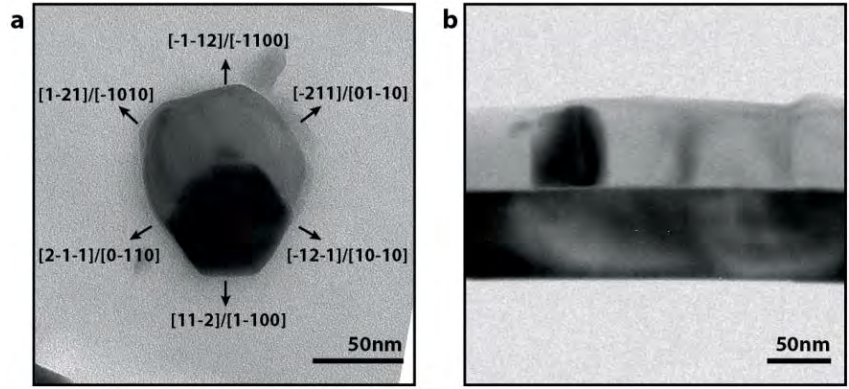

**Fig. S7. Grain boundary driven 'thick' three-facet shell of Al grown on the InAs NW sidefacets**. **a** Topview cross sectional TEM image along the NW axis, where the dominating faceting in the Al crystal follows the faceting of the InAs, as in the full shell example in Fig 4 in the main text. **b** Sideview TEM image, where the (11-2) out-of-plane variants in the Al alternate along the axis of the NW, which suggests that the growth was terminated in the grain boundary driven growth mode according to the discussion in S1 and S2.

## S7) Growth and characterization of an AlAs barrier between InAs and Al

Anisotropic shell growth (half shell) of lattice mismatched semiconductor materials in NWs, such as InAs-AlAs heterostructures, makes the wires bend strongly due to strain induced from the mismatch, and it is therefore difficult to grow uniform non-lattice matched 'half-shell' semiconductor wires. This also makes it impossible to grow an uniform metal phase on such structures. Thus, to overcome this problem a new type of growth scheme for III-V barrier segments is discussed, which we call *backward growth*, as illustrated in Fig. S8 a. First a thin layer of As is deposited on the InAs NW facets at low temperature, by shortly opening the As valve. After As deposition and when the background pressure in the growth chamber has reached below $10^{-10}$ torr, the shutter for the Al was opened, resulting in growth of AlAs (see the EDX linescan confirming that the presence of As extent into the barrier region in Fig. S8 b). As the AlAs/InAs interface has a lattice mismatch of ~7.5% (see Fig. S8 c), which is very close to the bulk mismatch, it indicates that the growth of AlAs nucleates away from the InAs interface. Moreover, the fact that these NW structures do not bend, even though they are half shell structures, is another indication that the NW is not strained at the interface and the growth does not start at the interface. Thus, nucleating from the As phase, the growth of AlAs stops when all the As is consumed and the unstrained AlAs meets the InAs. We speculate that the kinetic energy of the beam is enough to overcome the activation energy for nucleation AlAs, even though the substrate temperature is low. The shutter for the Al is kept open until the desired thickness of pure Al is reached. The advantage of this backward growth technique is that the hydrid structures not bend due to strain, and it is possible grow a uniform thickness of Al. The orientation of the Al phase, however, is not well defined and we have not observed a general preferred bicrystal orientation between AlAs and Al, but further optimization of this type of growth is needed to conclude on the properties and applications.

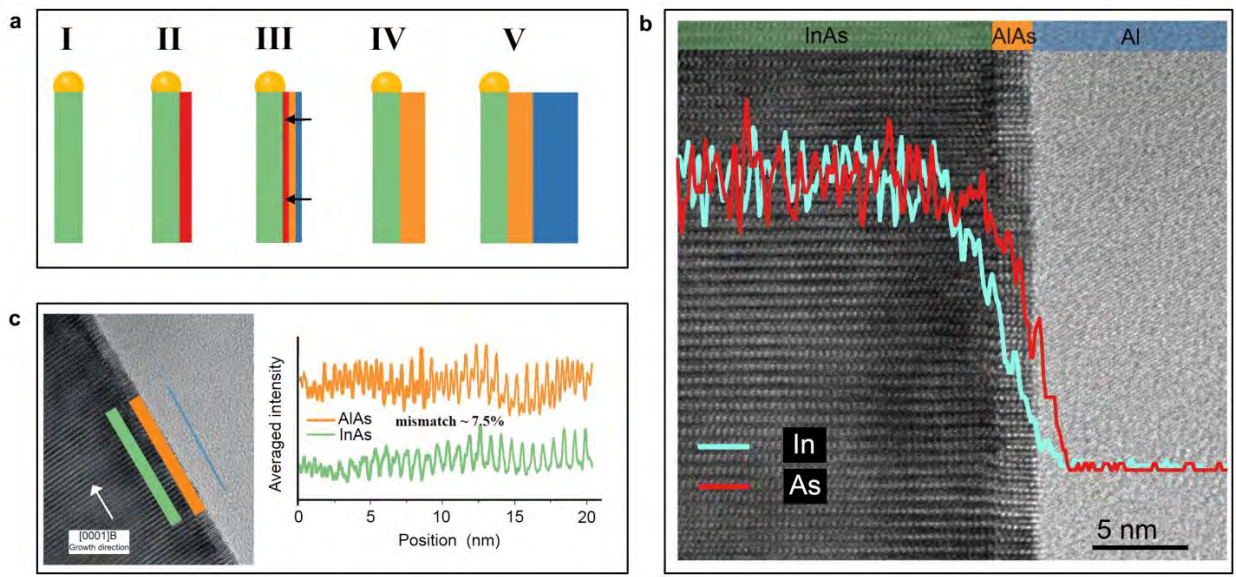

**Fig. S8 Backward barrier growth of AlAs in InAs/AlAs/Al hybrid NW crystals. a** Schematic illustration of the growth: **I** First the InAs NWs are grown via the Au catalysed VLS growth mechanism at T=420°C. **II** The substrate is cooled to T=-30°C and the As valve and shutter is opened for a short while and the As (red) is deposited on the NW facets. **III** When the pressure is back below $10^{-10}$ torr the Al shutter is opened and the Al (blue) promote nucleation of AlAs (orange in **III** and **IV**). **V** The Al flux is kept on until the desired thickness of Al (blue) is reached. **b** A TEM micrograph of the interface region, where three distinct regions are observed. An energy dispersive x-ray linescan intensity profile (a.u.) is over-layered across the interface, to show that the As extends into the barrier region due to the shift in the In and As profiles, indicating that it is AlAs. This is supported by measured averaged line intensity profiles along the NW length in **c** for several regions of both the InAs and the AlAs part, shows a lattice mismatch of 7.5%, which is close to the expected mismatch from relaxed bulk structures.

[1] Thompson, C. V., Floro, J. & H. I. Smith, Epitaxial grain growth in thin films. J. Appl. Phys **67** (9), 4099 (1990)

[2] Krogstrup, P. *et al.*, Advances in the theory of nanowire growth dynamics, J. Phys. D: Appl. Phys. **46** 313001 (2013)

[3] Carter, W.C., Roosen, A.R., Cahn, J.W. & Taylor, J.E., Shape Evolution by Surface Diffusion and Surface Attachment Limited Kinetics. Acta Metall. mater., **43**, 12, 43094323 (1995)

[4] Thomson, C. V., Grain growth in thin films, Annu. Rev. Mater. Sci. **20**, 245 (1990)

[5] Zheleva, T., Jagannadham, K. & Narayan, J. Epitaxial growth in large-lattice mismatch systems. J. Appl. Phys. **75,** 860 (1994)

[6] Novaco, A. D. & Mctague, J. P., Orientational Epitaxy – the orientational ordering of incommensurate structures. Phys. Rev. Lett. **38**, 22 (1977)

[7] Grey, F. & Bohr, J., A symmetry principle for epitaxial rotation, Europhys. Lett. **18**, 717-722 (1992)

[8] Pond, R. C. & Bollmann, W., The symmetry and Interfacial structure of bicrystals. Phil. Trans. R. Soc. Lond. A **292**, 449-472 (1979)

[9] Pond, R. C. & Vitek, V., Proc. R. Soc. Lond. A **357**, 1691 (1977)